\documentclass[aps,prd,superscriptaddress]{revtex4}

\newcommand{\ra}{\rightarrow}

\newcommand{\be}{\begin{equation}}
\newcommand{\ee}{\end{equation}}
\newcommand{\bea}{\begin{eqnarray}}
\newcommand{\eea}{\end{eqnarray}}

\newcommand{\s}{\smallskip}

\newcommand{\bc}{\begin{center}}
\newcommand{\ec}{\end{center}}
\newcommand{\bu}{\begin{underline}}
\newcommand{\eu}{\end{underline}}
\newcommand{\ty}{\textstyle}

\begin{document}
%\preprint{UCONN-94-6}
%\preprint{RU02-9-B}
\draft
\title{Gauge equivalence in QCD: the Weyl and Coulomb gauges.}
\author{Kurt Haller\thanks{E-mail: Kurt.Haller@uconn.edu}}
\affiliation{Department of Physics, University of Connecticut, Storrs, Connecticut
06269-3046} 
\author{Hai-cang Ren\thanks{E-mail}}
\affiliation{Physics Department, The Rockefeller University, 1230 York Avenue, New York, NY 10021-6399}
\begin{abstract}
The Weyl-gauge ($A_0^a=0)$ QCD Hamiltonian is unitarily transformed to a representation in 
which it is expressed entirely in terms of gauge-invariant quark and gluon fields. 
In a subspace of gauge-invariant states we have constructed that implement the non-Abelian Gauss's
law, this unitarily transformed Weyl-gauge Hamiltonian can be further transformed  and,
under appropriate circumstances, can be identified with the QCD Hamiltonian in the Coulomb gauge.
We demonstrate an isomorphism that materially facilitates the 
application of this Hamiltonian to a variety of physical processes,
including the evaluation of $S$-matrix elements. This 
isomorphism relates the gauge-invariant representation
of the Hamiltonian and the required set of gauge-invariant states 
to a Hamiltonian of the same functional 
form but dependent on ordinary unconstrained Weyl-gauge fields operating within a space of ``standard'' 
perturbative states. The fact that the gauge-invariant 
chromoelectric field is not hermitian  has important implications
for the functional form of the Hamiltonian finally obtained. When this nonhermiticity is taken into account, 
the ``extra'' vertices in Christ and Lee's Coulomb-gauge Hamiltonian are natural outgrowths of the formalism. When
this nonhermiticity is neglected, the Hamiltonian used in the earlier work of Gribov and others results. 
\end{abstract}
\narrowtext
\maketitle

\section{Introduction}
\label{sec:Intro}
In earlier work on QCD in the Weyl gauge ($A_0^a=0$), we have constructed 
gauge-invariant operator-valued quark and gluon fields;\cite{CBH2} 
these include the gauge-invariant quark field
\begin{equation}
{\psi}_{\sf GI}({\bf{r}})=V_{\cal{C}}({\bf{r}})\,\psi ({\bf{r}})
\;\;\;\mbox{\small and}\;\;\;
{\psi}_{\sf GI}^\dagger({\bf{r}})=
\psi^\dagger({\bf{r}})\,V_{\cal{C}}^{-1}({\bf{r}})\;,
\label{eq:psiqcdg1}
\end{equation}
where
\begin{equation}
V_{\cal{C}}({\bf{r}})=
\exp\left(\,-ig{\overline{{\cal{Y}}^\alpha}}({\bf{r}})
{\textstyle\frac{\lambda^\alpha}{2}}\,\right)\,
\exp\left(-ig{\cal X}^\beta({\bf{r}})
{\textstyle\frac{\lambda^\beta}{2}}\right)\;,
\label{eq:el1}
\end{equation}
\begin{equation}
V_{\cal{C}}^{-1}({\bf{r}})=
\exp\left(ig{\cal X}^\beta({\bf{r}})
{\textstyle\frac{\lambda^\beta}{2}}\right)\,
\exp\left(\,ig{\overline{{\cal{Y}}^\alpha}}({\bf{r}})
{\textstyle\frac{\lambda^\alpha}{2}}\,\right)\;,
\label{eq:eldagq1}
\end{equation}   
and where the $\lambda^a$ designate the Gell-Mann matrices. 
In these expressions ${\cal{X}}^\alpha({\bf{r}}) =
[\,{\textstyle\frac{\partial_j}{\partial^2}}A_j^\alpha({\bf{r}})]$, so that 
$\partial_i{\cal{X}}^\alpha({\bf{r}})$ is the $i$-th component of the longitudinal 
gauge field,~\cite{f1} and $\overline{{\cal Y}^{\alpha}}({\bf r})$ is defined as 
$\overline{{\cal Y}^{\alpha}}({\bf r})=[\,{\textstyle \frac{\partial_{j}}{\partial^{2}}
\overline{{\cal A}_{j}^{\alpha}}({\bf r})}]$. 
$\overline{{\cal A}_{j}^{\alpha}}({\bf r})$, which we 
refer to as the ``resolvent field'', is an operator-valued functional 
of the gauge field, and is represented in Refs.~\cite{CBH2} and \cite{HCC} as the 
solution of an integral equation. Constructing a gauge-invariant quark field by 
attaching $V_{\cal{C}}({\bf{r}})$ to the quark field 
$\psi$ represents an extension, into
the non-Abelian domain, of a method of creating gauge-invariant charged fields 
originated by Dirac for QED;~\cite{diracgauge} and, like Dirac's procedure,
this non-Abelian construction is free of path-dependent integrals.
An explicit demonstration that ${\psi}_{\sf GI}({\bf{r}})$
is invariant to non-Abelian gauge transformations has been given by implementing gauge transformations 
with the generator $\exp\{-i{\int}d{\bf y}{\hat {\cal G}}^{a}({\bf y})\omega^a({\bf y})\}$ where
${\hat {\cal G}}^{a}$ is the non-Abelian ``Gauss's law operator'' 
\be
{\hat {\cal G}}^{a}=\partial_i\Pi_i^a+gf^{abc}A_i^b\Pi_i^c+g\psi^{\dagger}({\bf r}){\textstyle
\frac{\lambda^a}{2}}\psi({\bf r})\,,
\label{eq:Ghat}
\ee
and $\omega^a$ is a number-valued gauge function.
With the use of this generator, under which  
\begin{equation}
{\psi}({\bf{r}})\,\rightarrow\,
\psi^\prime({\bf{r}})=\,
\exp\left(-i\omega^\alpha({\bf{r}})\,
{\textstyle\frac{\lambda^\alpha}{2}}\,\right)\,\psi({\bf{r}})\;
\label{eq:psitransf}
\end{equation}
and 
\be
A^b_i({\bf{r}}){\textstyle\frac{\lambda^b}{2}}\,\,\rightarrow\,\exp\left(-i\omega^\alpha({\bf{r}})
{\textstyle\frac{\lambda^\alpha}{2}}\right)
\left(A^b_i({\bf{r}}){\textstyle\frac{\lambda^b}{2}+\frac{i}{g}\partial_i}\right)
\exp\left(i\omega^\alpha({\bf{r}})
{\textstyle\frac{\lambda^\alpha}{2}}\,\right)\,,
\ee
it has been shown that 
$V_{\cal{C}}({\bf{r}})$ also gauge-transforms as 
\begin{equation}
V_{\cal{C}}({\bf{r}})\rightarrow 
V_{\cal{C}}({\bf{r}})\exp\left(i\omega^\alpha({\bf{r}})\,
{\textstyle\frac{\lambda^\alpha}{2}}\,\right)\;\;\;\;
\mbox{and}\;\;\;
V^{-1}_{\cal{C}}({\bf{r}})\rightarrow 
\exp\left(-i\omega^\alpha({\bf{r}})\,
{\textstyle\frac{\lambda^\alpha}{2}}\,\right)
V_{\cal{C}}^{-1}({\bf{r}})\;
\label{eq:Vgi}  
\end{equation}
so that ${\psi}_{\sf GI}({\bf{r}})$
remains gauge-invariant.\cite{CBH2} The resolvent field 
$\overline{{\cal{A}}^b_{j}}$ also has an important role 
in the gauge-invariant gauge field
\begin{equation}
{\sf A}_{{\sf GI}\,i}({\bf{r}})=[\,A_{{\sf GI}\,i}^{b}({\bf{r}})\,{\textstyle\frac{\lambda^b}{2}}\,]
=V_{\cal{C}}({\bf{r}})\,[\,A_{i}^b({\bf{r}})\,
{\textstyle\frac{\lambda^b}{2}}\,]\,
V_{\cal{C}}^{-1}({\bf{r}})
+{\textstyle\frac{i}{g}}\,V_{\cal{C}}({\bf{r}})\,
\partial_{i}V_{\cal{C}}^{-1}({\bf{r}})\;,
\label{eq:AdressedAxz}
\end{equation}
which can be shown to be the transverse field~\cite{CBH2}
\begin{equation}
A_{{\sf GI}\,i}^b({\bf{r}})=
A_{i}^{b\,{\sf T}}({\bf{r}}) +
\left[\delta_{ij}-{\textstyle\frac{\partial_{i}\partial_j}
{\partial^2}}\right]\overline{{\cal{A}}^b_{j}}({\bf{r}})=
\left[\delta_{ij}-{\textstyle\frac{\partial_{i}\partial_j}
{\partial^2}}\right]\left(A_{j}^{b}({\bf{r}})+\overline{{\cal{A}}^b_{j}}({\bf{r}})\right)\,.
\label{eq:Adressedthree1b}
\end{equation}
Eq. (\ref{eq:AdressedAxz}), as well as the fact that $A_{{\sf GI}\,i}^b({\bf{r}})$ 
and ${\hat {\cal G}}^c({\bf{x}})$ commute, demonstrate that $A_{{\sf GI}\,i}^b({\bf{r}})$  is 
gauge-invariant --- more precisely, invariant to ``small'' gauge transformations. 
We can also define a gauge-invariant chromoelectric field 
$E_{{\sf GI}\,i}^a=-\Pi_{{\sf GI}\,i}^a$.~\cite{HGrib} 
 A natural definition of $\Pi_{{\sf GI}\,i}^a$ in this formulation is
\be
\mbox{{\boldmath ${\Pi}$}}_{{\sf GI}\,i}({\bf{r}})=[\,{\Pi}_{{\sf GI}\,i}^{b}({\bf{r}})\,{\textstyle\frac{\lambda^b}{2}}\,]
=V_{\cal{C}}({\bf{r}})\,
{\textstyle\frac{\lambda^b}{2}}\,
V_{\cal{C}}^{-1}({\bf{r}})\,{\Pi}_{i}^b({\bf{r}})
\label{eq:Pigi1}
\ee  
or, equivalently,
\be
\Pi_{{\sf GI}\,i}^a=R_{ab}\Pi_i^b \;\;\;\mbox{where}\;\;\;R_{ab}={\textstyle\frac{1}{2}}{\sf Tr}
[\lambda^aV_{\cal{C}}\lambda^bV_{\cal{C}}^{-1}]\,,
\label{eq:Pigi2}
\ee
where $\Pi_i^a$ is the momentum conjugate to the gauge field $A_i^a$ in the Weyl gauge.
With the use of the commutator 
\be
\left[{\hat {{\cal G}}}^c({\bf{x}}),\,{\cal R}_{ab}({\bf{y}})\right]=
igf^{cbq}{\cal R}_{aq}({\bf{y}}){\delta}({\bf{x}}-{\bf{y}}),
\label{eq:gcommr}
\ee
obtained in Ref.~\cite{HGrib}, 
it is easy to verify that $\Pi_{{\sf GI}\,i}^a({\bf{y}})$ commutes with 
${\hat {{\cal G}}}^c({\bf{x}})$
and therefore also is gauge-invariant. \s

In this work we will use a representation, which we discuss in 
Section~\ref{sec:girep}, in which the Weyl-gauge QCD Hamiltonian 
is expressed entirely in terms of gauge-invariant fields. 
Since the gauge-invariant gauge field is transverse, it is of interest 
to relate this gauge-invariant formulation to the Coulomb gauge. We address this question in Section~\ref{subsec:Coul}. 
 In Section~\ref{sec:girep} we also show that the Weyl-gauge QCD Hamiltonian in this 
representation --- in which all operator-valued fields are gauge-invariant --- 
must be applied to a set of gauge-invariant states that
are solutions of the non-Abelian Gauss's law. In Section~\ref{sec:pert}, 
we address the problem that these states, which solve Gauss's law in QCD,
are complicated constructions that are difficult to use. We demonstrate an isomorphism in 
this section between this Hamiltonian, which operates on gauge-invariant  states,
and a corresponding Hamiltonian that is a functional of 
gauge-dependent Weyl-gauge fields and that operates on a set of ``standard'' perturbative states. Also, 
in Section~\ref{sec:pert}, we relate these Hamiltonians to those obtained from Coulomb-gauge formulations 
of QCD. We discuss the implications of our work in Section~\ref{sec:discuss}.
\s

\section{Relation of the gauge-invariant representation of the Weyl gauge to the Coulomb gauge.}
\label{sec:girep}
The QCD Hamiltonian in the Weyl gauge has been expressed in terms of gauge-invariant
operator-valued fields.~\cite{HGrib,BCH3} In this work, extensive use has been made of 
the unitary equivalence of ${\hat {\cal G}}^{a}$ --- the 
``Gauss's law operator'' given in Eq. (\ref{eq:Ghat}), which
imposes the non-Abelian Gauss's law --- to the ``pure glue'' version of that operator
\be
{\cal G}^{a}=\partial_i\Pi_i^a+gf^{abc}A_i^b\Pi_i^c
\label{eq:Gpg}
\ee 
as shown by 
\be
{\cal G}^{a}={\cal U_C}^{-1}{\hat {\cal G}^{a}}{\cal U_C}
\label{eq:UCtrans}
\ee
where 
\be
{\cal U_C}=\exp\left[i{\int}d{\bf r}{\cal X}^a({\bf{r}})j_0^a({\bf{r}})\right]
\exp\left[i{\int}d{\bf r}^\prime\,{\overline{{\cal{Y}}^c}}({\bf{r}}^\prime)j_0^c({\bf{r}}^\prime)\right]\,.
\label{eq:UCtrans2}
\ee
This unitary equivalence has been used to establish a new representation --- 
the ${\cal N}$ representation in which ${\cal G}^{a}$
represents the complete Gauss's law operator ${\hat {\cal G}}^{a}$, and $\psi$ represents the gauge-invariant
quark field because it commutes with  ${\cal G}^{a}$. The ${\cal N}$ 
representation  is unitarily equivalent to the  
${\cal C}$ representation in which ${\hat {\cal G}}^{a}$ and ${\psi}_{\sf GI}$ designate the Gauss's law 
operator and the gauge-invariant spinor (quark) field respectively. In the ${\cal N}$ representation, 
$j^a_0({\bf r})=g\psi^{\dagger}({\bf r}){\textstyle
\frac{\lambda^a}{2}}\psi({\bf r})$ and $j^a_i({\bf r})=g\psi^{\dagger}({\bf r}){\alpha}_i{\textstyle
\frac{\lambda^a}{2}}\psi({\bf r})$ are the gauge-invariant quark color charge
and quark color current densities respectively. \s

The Weyl-gauge QCD Hamiltonian can be transformed from its familiar ${\cal C}$-representation form 
\be
H=\int d{\bf r} \left\{\ {\textstyle \frac{1}{2}
\Pi^{a}_{i}({\bf r})\Pi^{a}_{i}({\bf r})
+ \frac{1}{4}} F_{ij}^{a}({\bf r}) F_{ij}^{a}({\bf r})+
{\psi^\dagger}({\bf r})
\left[\,\beta m-i\alpha_{i}
\left(\,\partial_{i}-igA_{i}^{a}({\bf r})
{\textstyle\frac{\lambda^\alpha}{2}}\,\right)\,\right]
\psi({\bf r})\right\}\,
\label{eq:HQCDC}
\ee
to the ${\cal N}$ representation, as shown by
\be
{\hat H}_{\sf GI}={\cal U_C}^{-1}H{\cal U_C}\,.
\label{eq:hunit}
\ee
This similarity transformation leaves the gauge field untransformed, but it transforms the quark field 
and the negative chromoelectric field as shown by~\cite{f2}
\be
{\cal U_C}^{-1}({\bf{x}})\,{\psi}({\bf{x}})\,{\cal U_C}({\bf{x}})=V_{\cal{C}}^{-1}({\bf{x}})\,\psi({\bf{x}})
\ee
and
\be
{\cal U_C}^{-1}({\bf{x}})\Pi_i^a({\bf{x}})\,{\cal U_C}({\bf{x}})=
\Pi_i^a({\bf{x}})-R_{ba}({\bf{x}}){\partial^{({\bf{x}})}_i}{\int}d{\bf y}{\cal D}^{\,bc}({\bf x},{\bf y})
j_0^c({\bf y})\,.
\ee
The transformed, ${\cal N}$-representation Hamiltonian ${\hat H}$ can be expressed entirely in terms of 
gauge-invariant variables by making use of the identities 
$$
R_{aq}R_{bq}=\delta_{ab},\;\;f^{duv}R_{ua}R_{vb}=f^{abq}R_{dq},\;\;\mbox{and}\;\;
\partial_iR_{ba}=-f^{uvb}R_{ua}P_{vi}\;\;\mbox{where}\;\;P_{vi}=-i{\sf Tr}[\lambda^vV_{\cal{C}}\,
\partial_{i}V_{\cal{C}}^{-1}]\,.
$$
%{from which we obtain}\;\;\;\Pi^a_i=R_{ba}\Pi_{{\sf GI}\,i}^b\,, 
The QCD Hamiltonian  in the ${\cal N}$ representation, expressed in terms of gauge-invariant fields, is
\bea
&&\!\!\!\!\!\!\!\!\!\!\!\!{\hat H}_{\sf GI}=\int d{\bf r}\left[ \ {\textstyle \frac{1}{2}}
\Pi^{a\,{\dagger}}_{{\sf GI}\,i}({\bf r})\Pi^{a}_{{\sf GI}\,i}({\bf r})
+  {\textstyle \frac{1}{4}} F_{{\sf GI}\,ij}^{a}({\bf r}) F_{{\sf GI}\,ij}^{a}({\bf r})-
{\psi^\dagger}({\bf r})\left(\beta m-i\alpha_{i}
\partial_{i}\right)\psi({\bf r})\right]+\nonumber \\
&&{\textstyle\frac{1}{2}}\int d{\bf x}d{\bf y}\left(J_{0\,({\sf GI})}^{a\,\dagger}({\bf x}) 
{\cal D}^{\,ab}({\bf x},{\bf y})j_0^b({\bf y})+j_0^b({\bf y})
\stackrel{\longleftarrow}{{\cal D}^{\,ba}}({\bf y},{\bf x})
J_{0\,({\sf GI})}^{a}({\bf x})\right)-\nonumber\\
&&{\textstyle\frac{1}{2}}{\int}d{\bf r}d{\bf x}d{\bf y}j_0^c({\bf y})
\stackrel{\longleftarrow}{{\cal D}^{\,ca}}({\bf y},{\bf r})
\partial^2{\cal D}^{\,ab}({\bf r},{\bf x})
j_0^b({\bf x})-{\int}d{\bf r}j^a_i({\bf r})A^a_{{\sf GI}\,i}({\bf{r}})+H_{\cal G}\,.
\label{eq:HQCDN}
\eea
where
\begin{equation}
F_{{\sf GI}\,ij}^{a}({\bf r})=\partial_jA_{{\sf GI}\,i}^{a}({\bf r})-\partial_iA_{{\sf GI}\,j}^{a}({\bf r})-
gf^{abc}A_{{\sf GI}\,i}^{b}({\bf r})A_{{\sf GI}\,j}^{c}({\bf r})\,,
\label{eq:FIJGI}
\end{equation}
from which it follows that 
\be
F_{{\sf GI}\,ij}^{a}({\bf r})=R_{aq}({\bf r})F_{ij}^{q}({\bf r})\,.
\label{eq:FijR}
\ee 
Because ${\hat H}_{\sf GI}$ is in the ${\cal N}$ representation, 
$\psi$ and ${\psi^\dagger}$ denote the gauge-invariant quark fields. ${\cal D}^{\,ab}({\bf x},{\bf y})$ 
is the inverse Faddeev-Popov operator, which we will discuss in Section~\ref{subsec:invFP}, and
$J_{0\,({\sf GI})}^{a}({\bf r})$ is the gauge-invariant gluon color charge density, defined as 
\be
 J_{0\,({\sf GI})}^{a}({\bf r})=g{f}^{abc}A^b_{{\sf GI}\,i}({\bf{r}})
{\Pi}_{{\sf GI}\,i}^c({\bf r})\,.
\label{eq:Jcolor}
\ee
Although ${\hat H}_{\sf GI}$ is hermitian, ${\Pi}_{{\sf GI}\,i}^a$ is not, 
because, as can be seen from Eq. (\ref{eq:Pigi2}),  
${\Pi}_{{\sf GI}\,i}^{a\,\dagger}={\Pi}_{i}^bR_{ab}$, and ${\Pi}_{i}^b$ does not commute with $R_{ab}$.
Similarly, $J_{0\,({\sf GI})}^{a}$ is not hermitian, and 
$J_{0\,({\sf GI})}^{a\,\dagger}=g{f}^{abc}{\Pi}_{{\sf GI}\,i}^{c\,\dagger}
A^b_{{\sf GI}\,i}\,.$  The last part of the QCD Hamiltonian is
\be
H_{\cal G}=-{\textstyle\frac{1}{2}}\int d{\bf x}d{\bf y}\big[{\cal G}_{\sf GI}^{a}({\bf x})
{\cal D}^{\,ab}({\bf x},{\bf y})j_0^b({\bf y})+
j_0^b({\bf y}){\cal D}^{\,ba}({\bf y},{\bf x}){\cal G}_{\sf GI}^{a}({\bf x})\big]
\label{eq:HamGauss}
\ee
where ${\cal G}_{\sf GI}^{a}$ is the gauge-invariant Gauss's law operator~\cite{HGrib}
$${\cal G}_{\sf GI}^{a}=\partial_i{\Pi}_{{\sf GI}\,i}^a+g{f}^{abc}A^b_{{\sf GI}\,i}
{\Pi}_{{\sf GI}\,i}^c=R_{ab}{\cal G}^{b}$$
which consists solely of gauge-invariant fields, every one of which commutes with ${\cal G}^{a}$, 
the Gauss's law operator in the ${\cal N}$ representation; ${\cal G}_{\sf GI}^{a}$ is hermitian 
because $R_{ab}$ and ${\cal G}^{b}$ commute.~\cite{HGrib}.\s

 Eq.~(\ref{eq:HQCDN}) resembles the QCD Hamiltonian in the 
Coulomb gauge.  The only direct interaction between color currents $j^a_i$ and the gauge field involve 
the transverse current only. The other interactions in which quarks participate are nonlocal, involve the  
quark color-charge density $j^a_0$, and are mediated by 
Green's functions that are the non-Abelian generalizations of the Abelian $\partial^{-2}$. These interactions
still involve the longitudinal component of the gauge-invariant
chromoelectric field, but we will show how this can be eliminated in Section \ref{subsec:Coul}.\s

\subsection{The inverse Faddeev-Popov operator.}
\label{subsec:invFP}
The Faddeev-Popov operator in the gauge-invariant representation of the Weyl gauge is 
\be
\partial{\cdot}D_{{({\bf x})}}^{ab}=\frac{\partial}{{\partial}x_i}
\left(\frac{\partial}{{\partial}x_i}\delta_{ab}
+gf^{a{q}b}A^{q}_{{\sf GI}\,i}({\bf x})\right)=\left(\frac{\partial}{{\partial}x_i}\delta_{ab}
+gf^{a{q}b}A^{q}_{{\sf GI}\,i}({\bf x})\right)\frac{\partial}{{\partial}x_i}\,;
\label{eq:tcom}
\ee
$\partial_i$ and $D_i$ commute because $A^{q}_{{\sf GI}\,i}$ is transverse. 
The Faddeev-Popov operator has a formal inverse, which can be represented as the series 
\be
{\cal D}^{\,bh}({\bf y},{\bf x})=\sum_{n=0}^{\infty}f_{(n)}^{{\vec \delta}bh}(-1)^{n+1}g^n\frac{1}{\partial^2}
\left({\cal T}_{(n)}^{\vec{\delta}}({\bf y})\delta({\bf y}-{\bf x})\right)\,,
\label{eq:FPinv}
\ee
where ${f}^{\vec{\alpha}bh}_{(n)}$ represents the chain of SU(3) structure constants
 \be
{f}^{\vec{{\alpha}}bh}_{(n)}={f}^{{\alpha}_1bs_1}\,\,{f}^{s_1{\alpha}_2s_2}\,{f}^{s_2{\alpha}_3s_3}\,\cdots\,
\,{f}^{s_{(n-2)}{\alpha}_{(n-1)}s_{(n-1)}}{f}^{s_{(n-1)}{\alpha}_nh}\,,
\label{eq:fproductN}
\ee
and where repeated superscripted indices are summed from $1{\rightarrow}8$;
for $n =1$; the chain reduces to ${f}^{\vec{{\alpha}}bh}_{1}={f}^{{{\alpha}}bh}$;
and for $n =0$, ${f}^{\vec{{\alpha}}bh}_{0}=-\delta_{bh}$. 
${\cal T}_{(n)}^{\vec{\alpha}}({\bf r})j_0^{h}({\bf{r}})$ is a special case of a general form 
${\cal T}_{(n)}^{\vec{\alpha}}({\bf r})\varphi^{h}({\bf{r}})$ for an arbitrary $\varphi^{h}({\bf{r}})$ given by 
 \begin{equation}
{\cal T}_{(n)}^{\vec{\alpha}}({\bf r})\varphi^{h}({\bf{r}})=
A_{{\sf GI}\,j(1)}^{{\alpha}(1)}({\bf r})\,
{\textstyle\frac{\partial_{j(1)}}{\partial^{2}}}
\left(A_{{\sf GI}\,j(2)}^{{\alpha}(2)}({\bf r})\,
{\textstyle\frac{\partial_{j(2)}}{\partial^{2}}}
\left(\cdots\left(A_{{\sf GI}\,j(n)}^{{\alpha}(n)}({\bf r})\,
{\textstyle\frac{\partial_{j(n)}}{\partial^{2}}}
\left(\varphi^{h}({\bf{r}})\right) \right)\right)\right)\,,
\label{eq:calTgi}
\end{equation}
with 
\be
{\cal T}_{(0)}^{\vec{\alpha}}({\bf r})\varphi^{h}({\bf{r}})=\varphi^{h}({\bf{r}})\;\;\mbox{and}\;\;
{\cal T}_{(1)}^{\vec{\alpha}}({\bf r})\varphi^{h}({\bf{r}})=A_{{\sf GI}\,i}^{{\alpha}}({\bf r})\,
{\textstyle\frac{\partial_{i}}{\partial^{2}}}\varphi^{h}({\bf{r}}). 
\label{eq:calTg01}
\ee
By expanding ${\cal D}^{\,bh}({\bf y},{\bf x})$ and combining terms of the same order in $g$,  it can be observed that,
as will be proven in Appendix~\ref{subsec:FPinv}, 
\be
{\partial}{\cdot}D^{ah}_{({\bf y})}\,{\cal D}^{\,hb}({\bf y},{\bf x})=\delta_{ab}\delta({\bf y}-{\bf x})
\label{eq:FPd} 
\ee
where $D^{ah}={\partial_i}\delta_{ah}+gf^{a{\gamma}h}A^\gamma_{{\sf GI}\,i}$ and that
\be
{\cal D}^{\,bh}({\bf y},{\bf x})\stackrel{\Longleftarrow}{\partial{\cdot}D^{ha}}_{({\bf x})}=
\delta_{ba}\delta({\bf y}-{\bf x})\,,
\label{eq:FPi}
\ee
where 
\be
\stackrel{\Longleftarrow}D_i^{hb}=
\left(\stackrel{\leftarrow}{\partial_i}\delta_{hb}-gf^{hqb}A^q_{{\sf GI}\,i}\right)
\label{eq:backarrowD}
\ee
and
\be
\stackrel{\Longleftarrow}{\partial{\cdot}D^{hb}}=
\left(\stackrel{\leftarrow}{\partial^2}\delta_{hb}-gf^{hqb}\stackrel{\leftarrow}{\partial_i}A^q_{{\sf GI}\,i}\right)
\label{eq:backarrow}
\ee
and the $\stackrel{\leftarrow}{}$ symbol 
indicates that  $\partial^2$ and $\partial_i$ differentiate to the left.
In demonstrating Eqs. (\ref{eq:FPd}) and (\ref{eq:FPi}), it can be helpful to use the expanded form of the 
$n$-th order term of the inverse Faddeev-Popov operator series 
%${\cal D}_{(n)}^{\,ah}({\bf y},{\bf x})$ which is given by    
\bea
{\cal D}_{(n)}^{\,ah}({\bf y},{\bf x})&&=g^nf^{\delta_1as_1}f^{s_1\delta_2s_2}{\cdots}f^{s_{(n-1)}\delta_nh}
{\int}\frac{d{\bf z}({\scriptstyle 1})}{4{\pi}|{\bf y}-
{\bf z}({\scriptstyle 1})|}A_{{\sf GI}\,l_1}^{{\delta}_1}
({\bf z}({\scriptstyle 1}))\frac{\partial}{{\partial}z({\scriptstyle 1})_{l_1}}
{\int}\frac{d{\bf z}({\scriptstyle 2})}
{4{\pi}|{\bf z}({\scriptstyle 1})-{\bf z}({\scriptstyle 2})|}
{\times}\nonumber \\
&&\!\!\!\!\!\!\!\!\!A_{{\sf GI}\,l_2}^{{\delta}_2}
({\bf z}({\scriptstyle 2}))\frac{\partial}{{\partial}z({\scriptstyle 2})_{l_2}}\;
{\cdots}{\int}\frac{d{\bf z}({\scriptstyle n})}{4{\pi}|{\bf z}({\scriptstyle n-1}))-
{\bf z}({\scriptstyle n})|}
A_{{\sf GI}\,l_n}^{{\delta}_n}
({\bf z}({\scriptstyle n}))\frac{\partial}{{\partial}z({\scriptstyle n})_{l_n}}
\frac{1}{{4{\pi}|{\bf z}({\scriptstyle n})-{\bf x}|}}
\label{eq:Pihermg}
\eea 
with 
\be
{\cal D}_{(0)}^{\,ah}({\bf y},{\bf x})=\frac{-\delta_{ah}}{4{\pi}|{\bf y}-{\bf x}|}
\label{eq:D0}
\ee
and
\be
{\cal D}_{(1)}^{\,ah}({\bf y},{\bf x})=gf^{\delta_1ah}
{\int}\frac{d{\bf z}}{4{\pi}|{\bf y}-
{\bf z}|}A_{{\sf GI}\,k}^{{\delta}_1}
({\bf z})\frac{\partial}{{\partial}z_k}\left(\frac{1}{4{\pi}|{\bf z}-{\bf x}|}\right)\,.
\label{eq:Da}
\ee
Integration by parts with respect to the ${\bf z}(i)$ and the identity ${f}^{\vec{{\alpha}}ah}_{(n)}=
(-1)^n{f}^{\vec{{\alpha}}ha}_{(n)}$ demonstrate that
\be
{\cal D}^{\,ah}({\bf y},{\bf x})={\cal D}^{\,ha}({\bf x},{\bf y}).
\label{eq:symmetry}
\ee
It is apparent from Eqs.~(\ref{eq:Pihermg})-(\ref{eq:Da}) that ${\cal D}^{\,bh}({\bf y},{\bf x})$ 
obeys the integral equation~\cite{fs}
\be
{\cal D}^{\,bh}({\bf y},{\bf x})=-\left(\frac{\delta_{bh}}{4{\pi}|{\bf y}-{\bf x}|}+gf^{{\delta}bs}
{\int}\frac{d{\bf z}}{4{\pi}|{\bf y}-{\bf z}|}A_{{\sf GI}\,k}^{{\delta}}
({\bf z})\frac{\partial}{{\partial}z_k}{\cal D}^{\,sh}({\bf z},{\bf x})\right)\,,
\label{eq:PIe}
\ee
which has these equations as an iterative solution.\s

Eq. (\ref{eq:FPinv}) enables us to express the commutator of the gauge-invariant 
gauge field and the negative gauge-invariant chromoelectric field as
\be
\left[\Pi^{b}_{{\sf GI}\,j}({\bf y})\,,A_{{\sf GI}\,i}^{a}({\bf{x}})\right]=-i\left(\delta_{ab}\delta_{ij}
\delta({\bf x}-{\bf y}))+\frac{\partial}{{\partial}y_j}{\cal D}^{\,bh}({\bf y},{\bf x})\stackrel{\Longleftarrow}
{D^{ha}_{i}}({\bf{x}})\right)\,.
\label{eq:PAcomm}
\ee
Eq.~(\ref{eq:PAcomm}) and the commutator, obtained in Ref.~\cite{HGrib}, 
\be
\left[{\Pi}_{{\sf GI}\,i}^{a}({\bf x})\,,{\Pi}_{{\sf GI}\,j}^{b}({\bf y})\right]=
ig\left\{\frac{\partial}{\partial x_i}{\cal D}^{{a}h}({\bf x},{\bf y}){f}^{hcb}
{\Pi}_{{\sf GI}\,j}^{c}({\bf y})-
\frac{\partial}{\partial y_j}{\cal D}^{{b}h}({\bf y},{\bf x}){f}^{hca}
{\Pi}_{{\sf GI}\,i}^{c}({\bf x})\right\}\,,
\label{eq:compsch}
\ee
are in agreement with those given by Schwinger for the Coulomb gauge,~\cite{schwingera} 
except for some differences in 
operator order. This fact suggests that the gauge-invariant Weyl-gauge field and the Coulomb-gauge field 
discussed by Schwinger are very similar. The differences in operator-order should be expected 
 because, in Ref.~\cite{schwingera},  ambiguities in operator order in the Coulomb gauge are resolved by 
symmetrizing noncommuting operator-valued quantities so that Coulomb-gauge operators are kept hermitian. In our 
work in the gauge-invariant formulation 
of the Weyl gauge, ambiguities in operator order do not arise.
When, because of a non-symmetric ordering of gauge fields and chromoelectric fields, 
 some gauge-invariant operator-valued quantities 
turn out not to be hermitian, we leave them that way in order to
avoid {\em ad hoc} changes in operator order.\s

Eq. (\ref{eq:compsch}) leads to the commutation rule for the 
transverse parts of $\Pi^{b}_{{\sf GI}\,j}({\bf y})$,~\cite{fa}
\be
\left[\Pi^{a\,{\sf T}}_{{\sf GI}\,i}({\bf x})\,,\Pi^{b\,{\sf T}}_{{\sf GI}\,j}({\bf y})\right]=0\,.
\label{eq:PPtr}
\ee
Eq. (\ref{eq:PAcomm}) leads to the 
commutator of the transverse part of $\Pi^{b}_{{\sf GI}\,j}({\bf y})$ 
and $A_{{\sf GI}\,i}^{a}({\bf{x}})$ (which is transverse)
\be
\left[\Pi^{b\,{\sf T}}_{{\sf GI}\,j}({\bf y})\,,A_{{\sf GI}\,i}^{a}({\bf{x}})\right]
=-i\delta_{ab}\left(\delta_{ij}-\frac{\partial_i\partial_j}{\partial^2}\right)\delta({\bf x}-{\bf y})\,;
\label{eq:PATcomm}
\ee 
Eq. (\ref{eq:PAcomm}) can be shown to be consistent with $\partial_iA_{{\sf GI}\,i}^{a}=0$ because 
\bea
\frac{\partial}{{\partial}x_i}\left[\Pi^{b}_{{\sf GI}\,j}({\bf y})\,,A_{{\sf GI}\,i}^{a}({\bf{x}})\right]&=&
-i\left(\delta_{ab}\delta_{ij}
\frac{\partial}{{\partial}x_j}\delta({\bf x}-{\bf y})+\frac{\partial}{{\partial}y_j}
{\cal D}^{\,bh}({\bf y},{\bf x})\stackrel{\Longleftarrow}
{\partial{\cdot}D^{ha}_{({\bf{x}})}}\right)\nonumber\\
&=&-i\delta_{ab}\delta_{ij}\left(
\frac{\partial}{{\partial}x_j}\delta({\bf x}-{\bf y})+\frac{\partial}{{\partial}y_j}\delta({\bf x}-{\bf y})\right)=0\,,
\label{eq:dcom}
\eea
and with $D_j\Pi^{b}_{{\sf GI}\,j}{\approx}0$ because
\be 
D^{bc}_j({\bf y})\left[\Pi^{c}_{{\sf GI}\,j}({\bf y})\,,A_{{\sf GI}\,i}^{a}({\bf{x}})\right]=0
\label{eq:DPcon}
\ee
trivially.  \s 

The Faddeev-Popov operator has a well-documented importance in non-Abelian gauge theories. Gribov 
has shown that gauge fields that
have been gauge-fixed to have a vanishing divergence can differ from each other,~\cite{gribov,gribovb}
and that the Faddeev-Popov operator does not have a unique inverse. 
In that same work, Gribov makes the suggestion that 
the zeros of the Faddeev-Popov operator 
$\partial^2{\delta}_{ac}+gf^{abc}A_i^b\partial_i$~might so intensify the
interaction between color charges that the effect could account for confinement.
Subsequent authors have reiterated this suggestion,~\cite{LMPR,cahill} and connections between the zeros 
of the Faddeev-Popov operator and color confinement have been discussed by other authors as 
well.~\cite{zwanz1,zwanz2,Cut}\s

Eqs. (\ref{eq:FPd}) and (\ref{eq:FPi}) are based on a series representation of 
the operator-valued ${\cal D}^{\,bh}({\bf y},{\bf x})$;
they are obtained by combining all terms of equal order in $g$ and noting cancellations within each order. They
do not, however, establish that ${\cal D}^{\,bh}({\bf y},{\bf x})$ is the 
unique inverse of the Faddeev-Popov operator. Questions about uniqueness can readily be formulated about
number-valued functions, but are very difficult to address for operator-valued quantities. 
Eqs. (\ref{eq:FPd}) and (\ref{eq:FPi}) establish that ${\cal D}^{\,bh}({\bf y},{\bf x})$
is an operator-valued inverse of ${\partial}{\cdot}D^{ah}_{({\bf y})}$ (acting on the left) and of
$\stackrel{\Longleftarrow}{\partial{\cdot}D^{ha}_{({\bf x})}}$ (acting on the right) without
addressing the question of its uniqueness. However, although $A_{{\sf GI}\,i}^{a}$
is an operator-valued quantity, the SU(2) versions of its constituents --- the Weyl-gauge field $A_{i}^a$ and the 
resolvent field $\overline{{\cal A}_{i}^{a}}$ --- 
can be, and often have been, represented by  number-valued realizations  as functions of spatial variables. 
Such realizations have been used extensively to study the topology of gauge fields.~\cite{gribov,gribovb,jackraj} When the 
integral equation for the resolvent field referred to in Section~\ref{sec:Intro} is expressed in terms of a number-valued 
hedgehog representation, it can be transformed into a nonlinear differential equation that was shown to have multiple 
solutions.~\cite{HCC} Moreover, this nonlinear differential equation
was shown to be very nearly identical in form to the one used by Gribov as a specific illustration of the fact that 
the Faddeev-Popov operator for the transverse SU(2) 
gauge field does not have a unique inverse. With this number-valued realization we 
were able to establish that the  gauge-invariant  field, which is transverse, has a 
Gribov ambiguity,~\cite{HCC} even though there are no Gribov copies 
of the  gauge-dependent Weyl-gauge field.~\cite{wein,bns,Singer} \s

In the context of the quantized theory --- for example, in 
${\hat H}_{\sf GI}$ --- we will represent ${\cal D}^{\,bh}({\bf y},{\bf x})$ as the operator-valued series
described in Eqs. (\ref{eq:FPinv}) and (\ref{eq:Pihermg}). Since each term in this series has 
unambiguous and self-consistent commutation relations with all other operator-valued quantities, the series
representation of ${\cal D}^{\,bh}({\bf y},{\bf x})$ is entirely satisfactory for 
determining the commutators of ${\hat H}_{\sf GI}$ with other gauge-invariant operators --- 
and therefore determining their time dependence --- even though  
number-valued realizations of the gauge-invariant gauge field lead to nonlinear 
integral equations that do not have unique solutions. \s

It may seem surprising that, starting in the Weyl gauge and expressing the QCD Hamiltonian in that gauge 
in terms of gauge-invariant variables can lead to a form of the Hamiltonian that, 
while never actually having been gauge-transformed,
has the same dynamical effect as the QCD Hamiltonian in the Coulomb gauge.  
But  a remarkably similar state of affairs obtains in QED. When QED is formulated in the temporal gauge, and 
a unitary transformation is carried out that is the Abelian analog of the one that leads to 
the Hamiltonian described in Eqs. (\ref{eq:HQCDN})-(\ref{eq:HamGauss}), the following 
result is obtained:\cite{khqedtemp,khelqed} The QED Hamiltonian in the temporal gauge, 
unitarily transformed by ${\exp}\left[i\int\frac{1}{\partial^2}\partial_iA_i({\bf r})j_0({\bf r})d{\bf r}\right]$ 
--- the Abelian analog of the transformation ${\cal U_C}$ described in Eq.~(\ref{eq:UCtrans2}) --- 
takes the form
\bea
{\hat H}_{QED}={\int}d{\bf r}&&\left[{\textstyle \frac{1}{2}}{\Pi}_i({\bf r}){\Pi}_i({\bf r})+
{\textstyle \frac{1}{4}}F_{ij}({\bf r})F_{ij}({\bf r})+
{\psi^\dagger}({\bf r})\left(\beta m-i\alpha_{i}\partial_{i}\right)\psi({\bf r})\right]\nonumber \\
-&&{\int}d{\bf r}\,j_i({\bf r})A^{\sf T}_i({\bf r})
+{\int}d{\bf r}d{\bf r}^{\prime} \frac{j_0({\bf r})
j_0({\bf r}^{\prime})}
{8{\pi}|{\bf r}-{\bf r}^{\prime}|}+H_g\,;
\label{eq:Hqedtc}
\eea
 $A^{\sf T}_i$ designates the transverse Abelian gauge field --- which, in Abelian theories,
is also the gauge-invariant field --- and $H_g$ can be expressed as 
\be
H_g=-{\textstyle \frac{1}{2}}{\int}d{\bf r}\left({\partial}_i{\Pi}_i({\bf r})\frac{1}{\partial^2}j_0({\bf r})+
j_0({\bf r})\frac{1}{\partial^2}{\partial}_i{\Pi}_i({\bf r})\right)\,.
\label{eq:Hg}
\ee
$H_g$ is the Abelian analog of $H_{\cal G}$, described in Eq.~(\ref{eq:HamGauss}).
The Abelian Gauss's law operator, ${\hat {\cal G}}={\partial}_i{\Pi}_i+j_0$, transforms into ${\partial}_i{\Pi}_i$
in the representation in which $\psi$ represents the gauge-invariant
electron field; and the states that implement Gauss's law, which originally 
are selected by ${\cal G}({\bf r})\,|{\Psi}({\bf r})\rangle=0$, are 
given by ${\partial}_i{\Pi}_i({\bf r})|\Phi\rangle=0$ in the transformed representation 
(or, as is more appropriate for Abelian gauge theories, by 
${\cal G}^{(+)}({\bf r})\,|{\Psi}\rangle=0$ and ${\partial}_i{\Pi}^{(+)}_i({\bf r})|\Phi\rangle=0$ respectively,
 where ${}^{(+)}$ designates the positive-frequency parts of operators).~\cite{gupbleu,khqedtemp}
As can be seen, ${\hat H}_{QED}$ also consists of two parts: the Hamiltonian for QED in the Coulomb gauge; and 
$H_g\,$, which has no effect on the time evolution of states that implement Gauss's law, but which ``remembers'' 
the fact that ${\hat H}_{QED}$ is the transformed Weyl-gauge Hamiltonian
by preserving the field equations for that gauge.
  An identical transformation applies to covariant-gauge QED, the sole difference 
being in the form of the $H_g$ produced by the transformation. \s

As we can see from Eqs.~(\ref{eq:HQCDN}), (\ref{eq:HamGauss}), (\ref{eq:Hqedtc}) and (\ref{eq:Hg}), 
and as will become even more evident  in Eq. (\ref{eq:Heff}), 
QCD and QED are strikingly similar in the relation between 
their Hamiltonians in different gauges when these are represented in terms of gauge-invariant fields. 
Nevertheless, there are important differences between QED and QCD in the significance of this relationship. 
One such difference is that, in QED, we may safely use the original untransformed Weyl-gauge or 
covariant-gauge Hamiltonian in a space of perturbative states when evaluating $S$-matrix elements, 
even though these gauge-dependent perturbative states fail to implement Gauss's law. This means that,
for perturbative calculations in QED, we can safely
use the Lagrangian
\be
{\cal L} = -{\textstyle \frac{1}{4}}
F_{\mu\nu}F^{\mu\nu} - j^\mu A_\mu + \bar{\psi}(i\gamma^\mu\partial_\mu - m)\psi+{\cal L}_g
\label{eq:L0}
\ee
with ${\cal L}_g=-A_0G$ for the Weyl gauge and 
${\cal L}_g=-G\partial_{\mu}A^\mu+\ty\frac{1}{2}(1-\gamma)G^2$ for the covariant gauge, 
without paying any attention to Gauss's law whatsoever.  
A corresponding practice in Weyl-gauge QCD is the use of the Weyl-gauge 
Hamiltonian $H$ in a Fock space of perturbative states that are
not annihilated by ${\cal G}^{a}$. There is, however, the following important difference 
between QED and QCD. The use of perturbative states in QED 
without implementing Gauss's law is permissible 
because, in QED, a unitary equivalence can be established between
$\partial_i\Pi_{i}$ and $\partial_i\Pi_{i}+j_0$, so that $\partial_i\Pi_{i}$ can be 
interpreted as $\partial_i\Pi_{i}+j_0$ in a new representation.~\cite{khqedtemp,khelqed} In this way, it can be 
shown that perturbative states that implement only $\partial_i\Pi_{i}({\bf r})\approx0$ 
instead of  $\partial_i\Pi_{i}+j_0({\bf r})\approx0$ may be used when evaluating $S$-matrix 
elements in QED; the only effect on $S$-matrix elements from this substitution consists of changes 
to the renormalization constants, which are unobservable.~\cite{hlren} But
this dispensation to ignore Gauss's law in perturbative calculations 
has not been shown to extend to QCD, because $D_i\Pi^a_{i}+j_0({\bf r})$ is 
unitarily equivalent only to $D_i\Pi^a_{i}$, but {\em not} to $\partial_i\Pi^a_{i}$; and states that implement the Gauss's
law $D_i\Pi^a_{i}\approx0$ cannot be perturbative states. In particular, the use of ${\hat H}_{\sf GI}$ for 
perturbative calculations using a space of perturbative states does not enjoy the same
protection that the corresponding practice has in QED.  In Section~\ref{sec:pert}, we will establish an 
isomorphism between the gauge-invariant states that implement the non-Abelian Gauss's law and perturbative 
states. This isomorphism enables us to substitute ``standard'' calculations with perturbative states for 
prohibitively difficult ones with gauge-invariant states. By this means, we provide for QCD a  
substitution rule, similar to the one available in QED, that permits the
use of perturbative Fock states in scattering calculations with the assurance that the results of these
calculations will agree with what would have been obtained if gauge-invariant operators and states had been
used.
\s

Another difference between QCD and QED is related to the fact that 
states that obey the condition 
\be
{\cal G}^a({\bf r})|\Psi\rangle=0
\label{eq:subconqcd}
\ee
are not normalizable. We can see this easily by
constructing, for example, the commutator of ${\cal G}^8({\bf r})$ and an integral operator
${\cal I}={\int}d{\bf r}^{\prime}A^8_j({\bf r}^\prime)\chi_j({\bf r}^\prime)$ where $\chi_j({\bf r}^\prime)$
is an arbitrary $c$-number-valued function. Since 
$$\left[\,{\cal I}\,,{\cal G}^8({\bf r})\right]=i\partial_i\chi_i({\bf r})\,,\;\;\;\mbox{and}\;\;\;
\langle\Psi|\left[\,{\cal I}\,,{\cal G}^8({\bf r})\right]|\Psi\rangle=
i\partial_i\chi_i({\bf r})\langle\Psi|\Psi\rangle\,,$$ and since ${\cal G}^8({\bf r})$ is hermitian so that
$\langle\Psi|{\cal G}^8({\bf r})=0$ as well as ${\cal G}^8({\bf r})|\Psi\rangle=0$, 
this leads to $\langle\Psi|\Psi\rangle=0$, in contradiction to the assumption that $|\Psi\rangle$ is normalizable. 
This argument is a simple extension of one that was applied to the Fermi 
subsidiary condition for QED.~\cite{fermi} In the case of QED, however, this difficulty can be remedied 
because the non-normalizability of the states that are annihilated by the Abelian Gauss's law operator 
is entirely caused by the unobservable longitudinal
nonpropagating photon ``ghost'' modes, which coincide exactly with the 
pure gauge degrees of freedom, and which can be
kept separate from the gauge-invariant transversely polarized propagating photons in a variety of ways. 
In QCD, however, transverse modes can be pure gauge, and 
we do not know of a similarly satisfactory resolution of the non-normalizability of 
the state vectors that satisfy Eq. (\ref{eq:subconqcd}).~\cite{f3,khqcdtemp} 
The previously-mentioned isomorphism, which will be demonstrated in Section~\ref{sec:pert}, mitigates this 
difficulty by establishing an equivalence between matrix elements evaluated with gauge-invariant states that
are not normalizable, and corresponding ones evaluated with perturbative states. \s

\subsection{Relation to QCD in the Coulomb gauge.}
\label{subsec:Coul}
Unlike the Weyl-gauge formulation of QCD, in which one
can simply set $A^a_0=0$ and impose canonical quantization rules on the remaining fields,~\cite{goldjack,jackiwt} 
the quantization of Coulomb-gauge QCD requires that constraints be explicitly taken into account. 
In constrained quantization --- one procedure for implementing consistency with constraints ---
this consistency is maintained by means of 
the so-called ``Dirac-brackets'', which replace the canonical equal-time commutation rules. 
When constrained quantization, such as the Dirac-Bergmann procedure,\cite{diracberg}
 is applied to the Coulomb gauge, the generator 
of infinitesimal gauge transformations becomes a constraint; it then must commute with all fields, which 
therefore are invariant to small gauge transformations. Under these 
circumstances, the gauge field would automatically be invariant to small gauge transformations, 
although it might have discrete numbers of gauge copies.\s
  
However, carrying out the constrained quantization of QCD in the Coulomb gauge is problematical;
one impediment stems from operator-ordering ambiguities of multilinear operator products. 
For example, in constrained quantization,  the matrix of
constraint commutators must be inverted. There are noncommuting operators in that matrix,
and it is at best problematical to keep track of operator order in the process of finding this inverse.
As a result, 
the Dirac brackets of some operators are not unambiguously specified.  
Because of the difficulties associated with the
quantization of QCD in the Coulomb gauge, a number of workers have avoided the direct quantization of 
Coulomb-gauge QCD, and have proceeded by treating the $A^a_0=0$ gauge fields 
as a set of Cartesian coordinates and the Coulomb-gauge fields as a set 
of curvilinear coordinates,
and have transformed from the former to the latter by using the familiar apparatus for 
such coordinate transformations.~\cite{christlee,lee,creutz,sakita} \s

In our work, we transform from the Weyl gauge to a representation in 
terms of gauge-invariant operator-valued fields. Our purpose is to implement gauge invariance,
not to carry out a gauge transformation. We do not impose transversality 
on the the gauge-invariant $A_{{\sf GI}\,i}^b$; in our work, $A_{{\sf GI}\,i}^b$ is transverse, but the transversality 
is not imposed as a condition --- it emerges as a consequence of its
gauge invariance. And the Gauss's law operator ${\cal G}^a$ 
does not vanish identically;  in our work, Gauss's law is a condition on a set of states 
(the implementation of Gauss's law by imposing it on a set of states is also discussed in
Refs.~\cite{christlee,creutz,schwingerb,rosstest}).\s

Because our formulation of QCD in terms of gauge-invariant fields differs significantly from those
whose purpose is to construct the QCD Hamiltonian in the Coulomb gauge, it is of interest 
to inquire how closely the resulting Hamiltonians resemble each other.     
In order to examine this question further, we will 
make some additional transformations of ${\hat H}_{\sf GI}$ that assume that the Hamiltonian acts only on states 
that implement Gauss's law. 
 When ${\hat H}_{\sf GI}$ appears in a matrix element between two states $|\Psi_\alpha\rangle$ 
and $\langle\Psi_\beta|$ that obey ${\cal G}^c({\bf{x}})\,|\Psi_\alpha\rangle=0$ and 
$\langle\Psi_\beta|{\cal G}^c({\bf{x}})=0$,
further transformations that eliminate the longitudinal component of ${\Pi}_{{\sf GI}\,i}^a$  are possible. 
For the case that ${\Pi}_{{\sf GI}\,i}^{c\,{\sf L}}$ appears adjacent to and directly to the left of 
such a state $|\Psi\rangle$, we can make the replacement
\be
{\Pi}_{{\sf GI}\,i}^{c\,{\sf L}}|\Psi\rangle=-{\textstyle \frac{\partial_i}{\partial^2}}J_{0\,({\sf GI})}^{c}|\Psi\rangle
\ee  
and, therefore, also 
\be
J_{0\,({\sf GI})}^{a}|\Psi\rangle=g{f}^{abc}A^b_{{\sf GI}\,i}\left({\Pi}_{{\sf GI}\,i}^{c\,{\sf T}}+
{\Pi}_{{\sf GI}\,i}^{c\,{\sf L}}\right)|\Psi\rangle=\left\{J_{0\,({\sf GI})}^{a\,{\sf T}}-g{f}^{abc}A^b_{{\sf GI}\,i}
{\textstyle \frac{\partial_i}{\partial^2}}J_{0\,({\sf GI})}^{c}\right\}|\Psi\rangle\,,
\label{eq:jjt}
\ee
where $J_{0\,({\sf GI})}^{a\,{\sf T}}$ is defined as 
$J_{0\,({\sf GI})}^{a\,{\sf T}}{\equiv}g{f}^{abc}A^b_{{\sf GI}\,i}{\Pi}_{{\sf GI}\,i}^{c\,{\sf T}}\,$.
Eq. (\ref{eq:jjt}) can be iterated, leading to
\be
J_{0\,({\sf GI})}^b\approx-\sum^\infty_{n=0}(-1)^ng^nf^{{\vec \alpha}bh}\left({\cal T}_{(n)}^{\vec{\alpha}}
J_{0\,({\sf GI}}^{h\,{\sf T}})\right)
\label{eq:Jexp}
\ee
where $\approx$ indicates that the replacement is valid only when the operators act on states $|\Psi\rangle$
that implement Gauss's law.
When $J_{0\,({\sf GI})}^{a\,\dagger}$ stands directly to the right of $\langle\Psi|$ states, we can 
similarly make the replacement 
\be
J_{0\,({\sf GI})}^{b\,\dagger}\approx-\sum^\infty_{n=0}(-1)^ng^nf^{{\vec \alpha}bh}\left({\cal T}_{(n)}^
{\vec{\alpha}}J_{0\,({\sf GI}}^{h\,{\sf T}})\right)^\dagger
\label{eq:Jexp_h}
\ee
where 
\be
\left\{{\cal T}_{(n)}^{\vec{\alpha}}({\bf r})J_{0\,({\sf GI})}^{h\,{\sf T}}({\bf r})\right\}^\dagger=
\left(\left(\left(\left(J_{0\,({\sf GI})}^{h\,{\sf T}\,\dagger}({\bf r})\right)\stackrel{\longleftarrow}
{{\textstyle \frac{\partial_{j(n)}}{\partial^{2}}}}
A_{{\sf GI}\,j(n)}^{{\alpha}(n)}({\bf r})\,\right)\cdots
\stackrel{\longleftarrow}{\textstyle\frac{\partial_{j(2)}}
{\partial^{2}}}A_{{\sf GI}\,j(2)}^{{\alpha}(2)}({\bf r})\,\right)
\stackrel{\longleftarrow}{\textstyle\frac{\partial_{j(1)}}{\partial^{2}}}
A_{{\sf GI}\,j(1)}^{{\alpha}(1)}({\bf r})\right)
\label{eq:TJdag}
\ee
and where the arrows indicate that differentiation is applied to the left. Similarly, 
$\Pi^{a}_{{\sf GI}\,i}({\bf r})$ and $\Pi^{a\,{\dagger}}_{{\sf GI}\,i}({\bf r})$
can be expressed as 
\be
\Pi^{b}_{{\sf GI}\,j}\approx{\Pi}_{{\sf GI}\,j}^{b\,{\sf T}}+
{\textstyle \frac{\partial_{j}}{\partial^{2}}}\left(\sum^\infty_{n=0}(-1)^ng^n
f^{{\vec \alpha}bh}\left({\cal T}_{(n)}^{\vec{\alpha}}
J_{0\,({\sf GI})}^{h\,{\sf T}}\right)\right)
\label{eq:Pirepa}
\ee
and 
\be
\Pi^{b\,{\dagger}}_{{\sf GI}\,j}\approx{\Pi}_{{\sf GI}\,j}^{b\,{\sf T\,\dagger}}+
{\textstyle \frac{\partial_{j}}{\partial^{2}}}\left(\sum^\infty_{n=0}(-1)^ng^n
f^{{\vec \alpha}bh}\left({\cal T}_{(n)}^{\vec{\alpha}}
J_{0\,({\sf GI})}^{h\,{\sf T}}\right)\right)^\dagger
\label{eq:Pirepb}
\ee
respectively.
We can combine Eqs. (\ref{eq:FPinv}) with  (\ref{eq:Jexp}) and (\ref{eq:Jexp_h}) to obtain 
\be
J_{0\,({\sf GI})}^b({\bf y})\approx\partial^2{\int}d{\bf x}{\cal D}^{\,ba}({\bf y},{\bf x})
J_{0\,({\sf GI})}^{a\,{\sf T}}({\bf x})
\label{eq:jfp}
\ee
and 
\be
J_{0\,({\sf GI})}^{b\,\dagger}({\bf y})\approx\partial^2{\int}d{\bf x}
J_{0\,({\sf GI})}^{a\,{\sf T}\,\dagger}({\bf x})
\stackrel{\longleftarrow}{{\cal D}^{\,ab}}({\bf x},{\bf y})\,;
\label{eq:jfpadj}
\ee
Eqs. (\ref{eq:Pirepa}) and (\ref{eq:Pirepb}) can be expressed as
\be
\Pi^{b}_{{\sf GI}\,j}({\bf y})\approx{\Pi}_{{\sf GI}\,j}^{b\,{\sf T}}({\bf y})-
\partial_j{\int}d{\bf x}{\cal D}^{\,ba}({\bf y},{\bf x})J_{0\,({\sf GI})}^{a\,{\sf T}}({\bf x})
\label{eq:Pirepc}
\ee
and 
\be
\Pi^{b\,\dagger\,}_{{\sf GI}\,j}({\bf y})\approx{\Pi}_{{\sf GI}\,j}^{b{\sf T}\,\dagger}({\bf y})-
\partial_j{\int}d{\bf x}J_{0\,({\sf GI})}^{a\,{\sf T}\,\dagger}({\bf x})
\stackrel{\longleftarrow}{{\cal D}^{\,ab}}({\bf x},{\bf y})
\label{eq:Pirepd}
\ee
respectively, where $J_{0\,({\sf GI})}^{a\,{\sf T}\,\dagger}({\bf x})$ represents the hermitian 
adjoint of $J_{0\,({\sf GI})}^{a\,{\sf T}}({\bf x})$.\s

We can define an ``effective'' Hamiltonian $({\hat H}_{\sf GI})_{\mbox{phys}}$, which is obtained by 
making the replacements 
described by Eqs.~(\ref{eq:jfp}) - (\ref{eq:Pirepd})) in ${\hat H}_{\sf GI}$ and 
excluding $H_{\cal G}$, since the latter will not contribute to any matrix elements in the 
physical space in which Gauss's law is implemented. With these replacements, we obtain
\bea
&&\!\!\!\!\!\!\!\!\!\!({\hat H}_{\sf GI})_{\mbox{phys}}=\int\!d{\bf r}\left[ \ {\textstyle \frac{1}{2}}
\Pi^{a\,{\sf T}\,\dagger}_{{\sf GI}\,i}({\bf r})\Pi^{a\,{\sf T}}_{{\sf GI}\,i}({\bf r})
+  {\textstyle \frac{1}{4}} F_{{\sf GI}\,ij}^a({\bf r}) F_{{\sf GI}\,ij}^{a}({\bf r})+
{\psi^\dagger}({\bf r})\left(\beta m-i\alpha_{i}
\partial_{i}\right)\psi({\bf r})\right]-{\int}d{\bf r}j^a_i({\bf r})A^a_{{\sf GI}\,i}({\bf{r}})\nonumber\\
&&\;\;\;\;\;\;\;\;\;\;\;\;-{\ty \frac{1}{2}}\int\!d{\bf r}d{\bf x}d{\bf y}\left(j_0^b({\bf x}))+
J_{0\,({\sf GI})}^{b\,{\sf T}\,\dagger}({\bf x})\right)
\stackrel{\longleftarrow}{{\cal D}^{\,ba}}({\bf x},{\bf r}){\partial^2}
{\cal D}^{\,ac}({\bf r},{\bf y})\left(j_0^c({\bf y}))+
J_{0\,({\sf GI})}^{c\,{\sf T}}({\bf y})\right)\,.
\label{eq:Heff}
\eea
$({\hat H}_{\sf GI})_{\mbox{phys}}$ is not identical to ${\hat H}_{\sf GI}$. But
$({\hat H}_{\sf GI})_{\mbox{phys}}$ can substitute for ${\hat H}_{\sf GI}$ as the 
generator of time-evolution when we embed the theory within a space of states  
$|\Psi_{\nu}\rangle$ that satisfy the non-Abelian Gauss's law,
${\cal G}^{a}({\bf{x}})|\Psi_\nu\rangle=0$. Because ${\cal G}^{a}({\bf{x}})$ is hermitian, the same 
state $|\Psi_\nu\rangle$ that obeys Eq. (\ref{eq:subconqcd}) also obeys 
${\langle}\Psi_\nu|{\cal G}^{a}({\bf{x}})=0$. Eq. (\ref{eq:HQCDN}) demonstrates
that when ${\hat H}_{\sf GI}$ appears in any ``allowed'' matrix element, $\Pi^{a}_{{\sf GI}\,i}$ and
$J_{0\,({\sf GI})}^{a}$ always are situated where they abut a 
``ket'' state vector $|\Psi_\alpha\rangle$ to their right; and $\Pi^{a\,{\dagger}}_{{\sf GI}\,i}$ 
and $J_{0\,({\sf GI})}^{a\,\dagger}$ always are situated 
where they abut a  ``bra'' state vector ${\langle}\Psi_\beta|$ to their
left. Since ${\hat H}_{\sf GI}$ will always be bracketed 
between two states ${\langle}\Psi_\beta|$ and $|\Psi_\alpha\rangle$ that implement Gauss's law, 
$\Pi^{a}_{{\sf GI}\,i}$ and $\Pi^{a\,{\dagger}}_{{\sf GI}\,i}$ can be replaced by their 
``soft'' equivalents shown in Eqs. (\ref{eq:Pirepc}) and (\ref{eq:Pirepd}) respectively; and 
$J_{0\,({\sf GI})}^b$ and $J_{0\,({\sf GI})}^{b\,\dagger}$ can similarly be replaced 
as shown in Eqs. (\ref{eq:jfp}) and (\ref{eq:jfpadj}) respectively. $({\hat H}_{\sf GI})_{\mbox{phys}}$
can therefore always be substituted for ${\hat H}_{\sf GI}$ in matrix elements, as long as attention
is paid to the need to restrict the space of state vectors to those that implement Gauss's law.  For example,
$\exp(-i{\hat H}_{\sf GI}t)|\Psi_{\alpha}\rangle$ can be 
replaced by $\exp\left(-i({\hat H}_{\sf GI})_{\mbox{phys}}t\right)|\Psi_{\alpha}\rangle$, 
since both will be required to project onto
states that implement Gauss's law, as shown by
\be
\exp(-i({\hat H}_{\sf GI})t)|\Psi_{\alpha}\rangle=
|\Psi_\nu\rangle\langle\Psi_\nu|\exp(-i({\hat H}_{\sf GI})t)|\Psi_{\alpha}\rangle\,,
\label{eq:timeco}
\ee
and
\bea
\langle\Psi_\nu|\exp(-i({\hat H}_{\sf GI})t)|\Psi_{\alpha}\rangle=&&\delta_{\nu\alpha}-it
\langle\Psi_\nu|{\hat H}_{\sf GI})|\Psi_{\alpha}\rangle+{\cdots}+\frac{(-it)^n}{n!}
\langle\Psi_\nu|{\hat H}_{\sf GI})|\Psi_{\mu_1}\rangle\,\times\nonumber \\
&&\langle\Psi_{\mu_1}|{\hat H}_{\sf GI})|\Psi_{\mu_2}\rangle
\langle\Psi_{\mu_2}|{\hat H}_{\sf GI})|\Psi_{\mu_3}\rangle
\cdots\langle\Psi_{\mu_{n-1}}|{\hat H}_{\sf GI})|\Psi_{\alpha}\rangle+\cdots\;.
\label{eq:timevo}
\eea
Each matrix element $\langle\Psi_{\mu_i}|{\hat H}_{\sf GI})|\Psi_{\mu_j}\rangle$  
in Eq. (\ref{eq:timevo}) can be replaced by 
$\langle\Psi_{\mu_i}|({\hat H}_{\sf GI})_{\mbox{phys}}|\Psi_{\mu_j}\rangle$, so that 
$\exp(-i({\hat H}_{\sf GI})t)|\Psi_{\alpha}\rangle$ can safely be replaced by 
$\exp\left(-i({\hat H}_{\sf GI})_{\mbox{phys}}t\right)|\Psi_{\alpha}\rangle$.
The time evolution imposed by 
${\hat H}_{\sf GI}$ on a state vector $|\Psi_{\alpha}\rangle$ for which 
${\cal G}^{c}({\bf{x}})|\Psi_\alpha\rangle=0$ takes place entirely within the space of states that
implement Gauss's law. In the case of a state vector $|\chi\rangle$ for which 
${\cal G}^{c}({\bf{x}})|\chi\rangle=|\chi^\prime\rangle$ where $|\chi^\prime\rangle$
is nonvanishing, 
\be
\langle\chi|\left[{\cal G}^{c}({\bf{x}})\,,\exp(-i{\hat H}_{\sf GI}t)\right]|\Psi_{\alpha}\rangle=
\langle\chi^\prime|\exp(-i{\hat H}_{\sf GI}t)|\Psi_{\alpha}\rangle=0
\label{eq:timexo}
\ee
because ${\cal G}^{c}({\bf{x}})$ and ${\hat H}_{\sf GI}$ commute. This requires the part of $\chi$
that fails to implement Gauss's law to be orthogonal to $\exp(-i{\hat H}_{\sf GI}t)|\Psi_{\alpha}\rangle$.
The only limitation on the validity of this argument is the non-normalizability of the states that 
implement Gauss's law, which complicates the algebraic properties of the 
$\{|\Psi_{\alpha}\rangle\}$ vector space. Nevertheless, Eqs.~(\ref{eq:timeco})-(\ref{eq:timexo})
show that we can restrict the space in which time evolution takes place
to state vectors that implement Gauss's law without 
compromising the unitarity of the time-evolved $|\Psi_{\alpha}(t)\rangle$ 
or of the $S$-matrix evaluated with such states. 
These considerations are also instrumental in allowing us to replace $\exp(-i{\hat H}_{\sf GI}t)$
with $\exp(-i({\hat H}_{\sf GI})_{\mbox{phys}}t)$.  ${\hat H}_{\sf GI}$ and $({\hat H}_{\sf GI})_{\mbox{phys}}$
both commute with ${\cal G}^a({\bf x})$ for all values of $a$, so that 
\be
{\cal G}^a({\bf x})\exp(-i({\hat H}_{\sf GI})_{\mbox{phys}}t)|\Psi_{\alpha}\rangle=
\exp(-i({\hat H}_{\sf GI})_{\mbox{phys}}t){\cal G}^a({\bf x})|\Psi_{\alpha}\rangle=0
\ee
as well as  
\be{\cal G}^a({\bf x})\exp(-i({\hat H}_{\sf GI})t)|\Psi_{\alpha}\rangle=
\exp(-i({\hat H}_{\sf GI})t){\cal G}^a({\bf x})|\Psi_{\alpha}\rangle=0\,.
\ee
The state vectors 
$\exp(-i({\hat H}_{\sf GI})t)|\Psi_{\alpha}\rangle$ and $\exp(-i({\hat H}_{\sf GI})_{\mbox{phys}}t)
|\Psi_{\alpha}\rangle$ therefore are gauge-invariant and implement Gauss's law just as 
$|\Psi_{\alpha}\rangle$ does.
\s

In comparing $({\hat H}_{\sf GI})_{\mbox{phys}}$ with expressions for the Coulomb-gauge Hamiltonian in the literature, 
we note that the only significant difference between $({\hat H}_{\sf GI})_{\mbox{phys}}$ and the Coulomb-gauge Hamiltonian
reported in Ref.~\cite{christlee} is that $\Pi_{{\sf GI}\,j}^{b\,{\sf T}\,\dagger}$, 
the hermitian adjoint of the transverse gauge-invariant chromoelectric field,
appears in  Eq. (\ref{eq:Heff}) where the expression ${\cal J}^{-1}\Pi_{{\sf GI}\,j}^{b\,{\sf T}}{\cal J}$ appears in  
Ref.~\cite{christlee}, where ${\cal J}=\det[\partial_i{\cdot}D_i]$. 
We will prove in Appendix~\ref{sec:Appb} that 
\be
\Pi_{{\sf GI}\,j}^{b\,{\sf T}\,\dagger}={\cal J}^{-1}\Pi_{{\sf GI}\,j}^{b\,{\sf T}}{\cal J}\,,
\label{eq:reneq2}
\ee
by using  Eq.~(\ref{eq:Pigi2})  and the identity
\be
\frac{\delta}{{\delta}A_i^q({\bf x})}\ln{\cal J}={\sf Tr}\left[\left(\partial{\cdot}D\right)^{-1}
\frac{\delta}{{\delta}A_i^q({\bf x})}\partial{\cdot}D\right]
\label{eq:reneq1}
\ee
 where the trace in Eq.~(\ref{eq:reneq1}) extends to the coordinates and the color 
indices.
With this demonstration, we see that Eq.~(\ref{eq:Heff}) and 
the Coulomb-gauge Hamiltonian described in Eq. (4.65) in Ref.~\cite{christlee} are identical.
It is also of interest to compare Eq.~(\ref{eq:Heff})  with 
the Coulomb-gauge Hamiltonian in Ref.~\cite{gribov} as well as in the work of a number of other 
authors who used the same form of the Hamiltonian. The Hamiltonian in Ref.~\cite{gribov} differs from the Hamiltonian 
described by Eq. (4.65) in Ref.~\cite{christlee} in the fact that $\Pi_{{\sf GI}\,j}^{b\,{\sf T}}$
rather than $\Pi_{{\sf GI}\,j}^{b\,{\sf T}\,\dagger}$
appears in Ref.~\cite{gribov} in place of ${\cal J}^{-1}\Pi_{{\sf GI}\,j}^{b\,{\sf T}}{\cal J}$
in Ref.~\cite{christlee}; there is also the trivial difference that Ref.~\cite{gribov} deals with ``pure glue'' 
QCD so that the quark field is not included. \s

This discrepancy raises the question  of the hermiticity of the operator-valued 
transverse gauge-invariant chromoelectric field 
$\Pi_{{\sf GI}\,j}^{b\,{\sf T}}$, which is of considerable importance for determining the dynamical 
effects of $({\hat H}_{\sf GI})_{\mbox{phys}}$.
One way of addressing this question is to use Eq.~(\ref{eq:Pigi2}) and Eq.~(65) in Ref.~\cite{HGrib} to obtain
\be
\Pi_{{\sf GI}\,j}^{b\,\dagger}({\bf y})-\Pi_{{\sf GI}\,j}^b({\bf y})=
\left[\Pi_j^q({\bf y})\,,R_{bq}({\bf y})\right]=igf^{hcb}
{\frac{\partial}{{\partial}y_j}}{\cal D}^{\,ch}({\bf y},{\bf y})
\label{eq:Tcomm} 
\ee
where the partial derivative acts on only the {\em first} ${\bf y}$ argument in ${\cal D}^{\,ch}({\bf y},{\bf y})$.
We might have expected that the transverse parts of $\Pi_{{\sf GI}\,j}^{b\,\dagger}({\bf y})$ and 
$\Pi_{{\sf GI}\,j}^b({\bf y})$ would be identical since  any functionals of the form 
$(\delta_{i,j}-\frac{\partial_i\partial_j}{\partial^2}){\textstyle \partial_j}{\xi}({\bf y})$ would 
necessarily vanish. Such a conclusion would not, however,  be correct in this case, because in
${\textstyle \frac{\partial}{{\partial}y_j}}{\cal D}^{\,qh}({\bf y},{\bf y})$, the partial
derivative differentiates only the {\em first} ${\bf y}$ in ${\cal D}^{\,qh}({\bf y},{\bf y})$.
We can make use of Eq.~(\ref{eq:PIe}) and the fact that $f^{hcb}\delta_{hc}=0$ to express Eq.~(\ref{eq:Tcomm})
as 
\be
\Pi_{{\sf GI}\,j}^{b\,\dagger}({\bf y})-\Pi_{{\sf GI}\,j}^b({\bf y})=
ig^2f^{hcb}f^{{\delta}cs}
{\int}{d{\bf z}}{ \frac{\partial}{{\partial}y_j}}\left(\frac{1}{4{\pi}|{\bf y}-{\bf z}|}\right)
A_{{\sf GI}\,k}^{{\delta}}({\bf z})\frac{\partial}{{\partial}z_k}{\cal D}^{\,sh}({\bf z},{\bf y}) 
\label{eq:pipiherm}
\ee
and we can extract the transverse parts to obtain
\be
\Pi_{{\sf GI}\,j}^{b\,{\sf T}\,\dagger}({\bf y})-\Pi_{{\sf GI}\,j}^{b\,{\sf T}}({\bf y})=
ig^2f^{hcb}f^{{\delta}cs}\left(\delta_{j,\ell}-\frac{\partial^{({\bf y})}_j
\partial^{({\bf y})}_\ell}{\partial^2}\right)
{\int}{d{\bf z}}{ \frac{\partial}{{\partial}y_\ell}}\left(\frac{1}{4{\pi}|{\bf y}-{\bf z}|}\right)
A_{{\sf GI}\,k}^{{\delta}}({\bf z})\frac{\partial}{{\partial}z_k}{\cal D}^{\,sh}({\bf z},{\bf y})\,. 
\label{eq:pipiTherm}
\ee
Eq.~(\ref{eq:pipiTherm}) makes it clear that $\Pi_{{\sf GI}\,j}^{b\,{\sf T}\,\dagger}({\bf y})-
\Pi_{{\sf GI}\,j}^{b\,{\sf T}}({\bf y})$ is not the transverse projection of a gradient 
and therefore cannot be presumed to vanish.\s  

Equally compelling evidence that  
$\Pi_{{\sf GI}\,j}^{b\,{\sf T}}$ is not identical to its hermitian adjoint is provided by the observation that 
the commutators $\left[\Pi_{{\sf GI}\,i}^{a\,{\sf T}}({\bf x})\,,\Pi_{{\sf GI}\,j}^{b\,{\sf T}\,\dagger}({\bf
y})\right]$ and $\left[\Pi_{{\sf GI}\,i}^{a\,{\sf T}}({\bf x})\,,\Pi_{{\sf GI}\,j}^{b\,{\sf T}}({\bf y})\right]$
differ. The  latter vanishes, as is shown by Eqs.~(\ref{eq:compsch})-(\ref{eq:PPtr}). However, use of
Eq.~(\ref{eq:Pigi2})  and the commutation rules for the underlying Weyl-gauge fields lead to 
\be
\left[\Pi_{{\sf GI}\,i}^{a\,{\sf T}}({\bf x})\,,\Pi_{{\sf GI}\,j}^{b\,{\sf T}\,\dagger}({\bf y})\right]=
g^2f^{hca}f^{pdb}\left(\delta_{ik}-{\textstyle \frac{\partial^{({\bf x})}_i\partial^{({\bf x})}_k}{\partial^2}}\right)
\left(\delta_{jl}-{\textstyle \frac{\partial^{({\bf y})}_j\partial^{({\bf y})}_l}{\partial^2}}\right)
{\textstyle \frac{\partial}{{\partial}y_l}}{\cal D}^{\,dh}({\bf y},{\bf x})
{\textstyle \frac{\partial}{{\partial}x_k}}{\cal D}^{\,cp}({\bf x},{\bf y})\,,
\label{eq:anomcom}
\ee
and an alternate derivation based on Eqs.~(\ref{eq:PATcomm}) and (\ref{eq:Tcomm}) confirms that result.
Similarly to what we observed in connection with Eq.~(\ref{eq:Tcomm}), the derivatives 
$\frac{\partial}{{\partial}y_j}$ and $\frac{\partial}{{\partial}x_i}$ each differentiate part, but not all
of the ${\bf y}$ and ${\bf x}$ dependence, respectively, of the product ${\cal D}^{\,dh}({\bf y},{\bf x})
{\cal D}^{\,cp}({\bf x},{\bf y})$ in Eq.~(\ref{eq:anomcom}). The transverse projections of 
${\textstyle \frac{\partial}{{\partial}y_l}}{\cal D}^{\,dh}({\bf y},{\bf x})
{\textstyle \frac{\partial}{{\partial}x_k}}{\cal D}^{\,cp}({\bf x},{\bf y})$ therefore will not vanish, and 
 $\left[\Pi_{{\sf GI}\,i}^{a\,{\sf T}}({\bf x})\,,\Pi_{{\sf GI}\,j}^{b\,{\sf T}\,\dagger}({\bf y})\right]\neq0$.
Since $\left[\Pi^{a\,{\sf T}}_{{\sf GI}\,i}({\bf x})\,,\Pi^{b\,{\sf T}}_{{\sf GI}\,j}({\bf y})\right]$ and 
$\left[\Pi_{{\sf GI}\,i}^{a\,{\sf T}}({\bf x})\,,\Pi_{{\sf GI}\,j}^{b\,{\sf T}\,\dagger}({\bf y})\right]$ differ
$\Pi_{{\sf GI}\,j}^{b\,{\sf T}}({\bf y})$ and $\Pi_{{\sf GI}\,j}^{b\,{\sf T}\,\dagger}({\bf y})$
cannot be identical.\s

\section{Isomorphism and its implication for the scattering amplitude.}
\label{sec:pert}
In the preceding sections we have obtained a description of QCD that took the Weyl-gauge formulation
as its point of departure, and arrived at a Hamiltonian in which all operator-valued fields --- the gauge 
field, the chromoelectric field, as well as the quark field --- are gauge-invariant, and only 
the transverse components of the chromoelectric fields appear in the Hamiltonian 
$({\hat H}_{\sf GI})_{\mbox{phys}}$. It was necessary, however, to restrict use of this 
Hamiltonian to a space in which all state vectors implement the non-Abelian Gauss's law; and these state vectors 
are complicated constructions that are not easy to use. In this section, we will show how isomorphisms can
be established that enable us to identify $({\hat H}_{\sf GI})_{\mbox{phys}}$ with a Hamiltonian that 
can be used in a space of ordinary, conventional perturbative states.\s 

To review the relation between gauge-invariant and perturbative states: In Ref.~\cite{CBH2}, a set of 
states was constructed in the form 
\be
|\Psi_i\rangle=\Psi|\phi_i\rangle
\label{eq:psistate} 
\ee
where the operator-valued $\Psi$ was given as 
\be
\Psi=\sum_{n=0}\frac{i^n}{n!}\Psi_n
\label{eq:psia}
\ee
with
\be
\Psi_n=\int{d{\bf r}_1}\cdots{d{\bf r}_n}\,\overline{{\cal{A}}^{q(1)}_{k(1)}}({\bf{r}}_1)\cdots
\overline{{\cal{A}}^{q(n)}_{k(n)}}({\bf{r}}_n)\Pi^{q(1)}_{k(1)}({\bf{r}}_1)\cdots\Pi^{q(n)}_{k(n)}({\bf{r}}_n)\,.
\label{eq:psib}
\ee
$|\phi_i\rangle$ designates one of a set of states that is annihilated by 
$\partial_j\Pi^{b}_{j}$. These $|\phi_i\rangle$ states --- 
the so-called ``Fermi'' states --- are related to ``standard'' perturbative states $|{\sf p}_i\rangle$ by 
\be
|\phi_i\rangle=\Xi|{\sf p}_i\rangle\,;
\label{eq:ferma}
\ee
$\Xi$ was given in Ref.~\cite{bch1}, where it was also shown that  
$\partial_j\Pi^{b}_{j}({\bf{r}})\,\Xi|{\sf p}_i\rangle=0$,  where $|{\sf p}_i\rangle$
designates one of a set of ``standard'' perturbative states annihilated by all annihilation operators
for fermion and transverse gauge field excitations. This set of
perturbative states will be described more fully later in this section, and will turn out to be identical to
perturbative states in QED, except for the fact that the gluon operators carry a Lie group index, while the photons do
not. Since $\partial_j\Pi^{b}_{j}$ annihilates
any $|\phi_i\rangle$ state,  we can see that, in $|\Psi_i\rangle$ states, the negative chromoelectric field 
$\Pi^{q(\ell)}_{k(\ell)}({\bf{r}}_\ell)$ in $\Psi$ can be replaced by its transverse part 
$\Pi^{q(\ell)\,{\sf T}}_{k(\ell)}({\bf{r}}_\ell)$, because the longitudinal parts vanish when acting on a 
$|\phi_i\rangle$ state. Furthermore, in Eq.~(\ref{eq:psib}), every transverse  
$\Pi^{q(\ell)\,{\sf T}}_{k(\ell)}({\bf{r}}_\ell)$ is integrated with an 
$\overline{{\cal{A}}^{q(\ell)}_{k(\ell)}}({\bf{r}}_\ell)$ in each variable, ${\bf r}_\ell$,  
and only the transverse components $\overline{{\cal{A}}^{q(\ell)\,{\sf T}}_{k(\ell)}}({\bf{r}}_\ell)$ 
will survive this integration in the $|\Psi_i\rangle$ states, 
which become
\be
|\Psi_i\rangle=\sum_{n=1}\frac{i^n}{n!}\int{d{\bf r}_1}\cdots{d{\bf r}_n}\,
\overline{{\cal{A}}^{q(1)\,{\sf T}}_{k(1)}}({\bf{r}}_1)\cdots
\overline{{\cal{A}}^{q(n)\,{\sf T}}_{k(n)}}({\bf{r}}_n)\Pi^{q(1)\,{\sf T}}_{k(1)}({\bf{r}}_1)\cdots
\Pi^{q(n)\,{\sf T}}_{k(n)}({\bf{r}}_n)|\phi_i\rangle\,.
\label{eq:psic}
\ee
In Ref.~\cite{CBH2}, it was shown that 
\be
A_{{\sf GI}\,j}^a({\bf r})\Psi\,|\phi_i\rangle=\Psi\,A_{j}^{a\,{\sf T}}({\bf r})|\phi_i\rangle\,.
\label{eq:ra}
\ee
In Appendix~\ref{sec:Appd}, we will use Eq.~(\ref{eq:psic}) to show that 
\be
\Pi_{{\sf GI}\,j}^{c\,{\sf T}}({\bf r})\Psi\,|\phi_i\rangle=
\Psi\,\Pi_{j}^{c\,{\sf T}}({\bf r})|\phi_i\rangle\,.
\label{eq:rb}
\ee
Since the Hamiltonian $({\hat { H}}_{\sf GI})_{\mbox{phys}}$ consists of transverse fields only, 
Eqs.~(\ref{eq:ra}) and (\ref{eq:rb}) afford us an opportunity to shift $({\hat { H}}_{\sf GI})_{\mbox{phys}}$
from the left-hand side of $\Psi$ to the right, with a concomitant substitution  of 
transverse Weyl-gauge fields for the corresponding gauge-invariant fields. The one impediment to this 
process is that $\Pi^{b\,{\sf T}\,\dagger}_{{\sf GI}\,j}$, the hermitian adjoint of 
$\Pi_{{\sf GI}\,j}^{b\,{\sf T}}$, also appears in $({\hat { H}}_{\sf GI})_{\mbox{phys}}$, and
Eq.~(\ref{eq:rb}) only applies to $\Pi_{{\sf GI}\,j}^{b\,{\sf T}}$ and not to $\Pi^{b\,{\sf
T}\,\dagger}_{{\sf GI}\,j}$. To remove that  impediment, we use Eq.~(\ref{eq:reneq2}) to substitute
${\cal J}^{-1}\Pi_{{\sf GI}\,j}^{b\,{\sf T}}{\cal J}$  for $\Pi_{{\sf GI}\,j}^{b\,{\sf
T}\,\dagger}$, and express $({\hat { H}}_{\sf GI})_{\mbox{phys}}$ as
\bea
\!\!\!\!\!\!\!\!\!&&({\hat { H}}_{\sf GI})_{\mbox{phys}}=\int\!d{\bf r}\left\{  {\textstyle \frac{1}{2}}
{\cal J}^{-1}\Pi_{{\sf GI}\,j}^{b\,{\sf T}}({\bf r}){\cal J}\Pi^{a\,{\sf T}}_{{\sf GI}\,i}({\bf r})
+  {\textstyle \frac{1}{4}} {F}_{{\sf GI}\,ij}^a({\bf r}) {F}_{{\sf GI}\,ij}^{a}({\bf r})+
{\psi^\dagger}({\bf r})\left(\beta m-i\alpha_{i}
\partial_{i}\right)\psi({\bf r})\right.-\nonumber\\
&&\left.j^a_i({\bf r})A^a_{{\sf GI}\,i}({\bf{r}})\right\}-
{\ty \frac{1}{2}}\int\!d{\bf r}d{\bf x}d{\bf y}\left(j_0^b({\bf x})+
J_{0\,({\sf GI})}^{b\,{\sf T}\,\dagger}({\bf x})\right)
\stackrel{\longleftarrow}{{\cal D}^{\,ba}}({\bf x},{\bf r}){\partial^2}
{\cal D}^{\,ac}({\bf r},{\bf y})\left(j_0^c({\bf y})+
J_{0\,({\sf GI})}^{c\,{\sf T}}({\bf y})\right)\nonumber \\
\label{eq:h1}
\eea
with
$$J_{0\,({\sf GI})}^{a\,{\sf T}\,\dagger}=gf^{abc}{\cal J}^{-1}\Pi_{{\sf GI}\,i}^{c\,{\sf T}}{\cal J}
A^b_{{\sf GI}\,i}\,.$$
We can define a ``hermitized'' transverse gauge-invariant negative chromoelectric field ${\cal P}^{b\,{\sf T}}_{j}$
\be
{\cal P}^{b\,{\sf T}}_{j}({\bf r})={\cal J}^{-\frac{1}{2}}\Pi_{{\sf GI}\,j}^{b\,{\sf T}}({\bf r})
{\cal J}^{\frac{1}{2}}\,.
\label{eq:ppi}
\ee
As can be seen from Eq.~(\ref{eq:reneq2}), ${\cal P}^{b\,{\sf T}}_{j}$ is hermitian, since
\be
{\cal P}^{b\,{\sf T}\,\dagger}_{j}({\bf r})={\cal J}^{\frac{1}{2}}\Pi_{{\sf GI}\,j}^{b\,{\sf T}\,\dagger}({\bf r})
{\cal J}^{-\frac{1}{2}}={\cal J}^{-\frac{1}{2}}\left({\cal J}\Pi_{{\sf GI}\,j}^{b\,{\sf T}\,\dagger}({\bf r})
{\cal J}^{-1}\right){\cal J}^{\frac{1}{2}}={\cal P}^{b\,{\sf T}}_{j}({\bf r})\,.
\label{eq:hermpa}
\ee
An important consideration for this argument is the fact that ${\cal J}^{\frac{1}{2}}$ is hermitian, which is proven in 
Appendix~\ref{sec:Appc}. In the same appendix, we also prove that the canonical commutation relations between 
$\Pi_j^{b{\sf T}}$'s and $A_{{\sf GI}j}^a$'s and that among $\Pi_j^{b{\sf T}}$'s remain unmodified with $\Pi_j^{b{\sf T}}$'s
replaced by ${\cal P}_{{\sf GI}j}^{b{\sf T}}$'s. 
We then find that
\be
\Pi_{{\sf GI}\,j}^{b\,{\sf T}}({\bf r})={\cal J}^{\frac{1}{2}}{\cal P}^{b\,{\sf T}}_{j}({\bf r})
{\cal J}^{-{\frac{1}{2}}}\;\;\mbox{and}\;\;\Pi_{{\sf GI}\,j}^{b\,{\sf T}\,\dagger}({\bf r})=
{\cal J}^{-\frac{1}{2}}{\cal P}^{b\,{\sf T}}_{j}({\bf r})
{\cal J}^{{\frac{1}{2}}}\,.
\label{eq:hermpb}
\ee 
Eq.~(\ref{eq:hermpb}) transforms from the nonhermitian $\Pi_{{\sf GI}\,j}^{b\,{\sf T}}$ 
and $\Pi_{{\sf GI}\,j}^{b\,{\sf T}\,\dagger}$ to the 
hermitian ${\cal P}^{b\,{\sf T}}_{j}$ (not, however, unitarily, since ${\cal J}^{\frac{1}{2}}$ 
is hermitian and not the hermitian adjoint of ${\cal J}^{-{\frac{1}{2}}}$). 
Transformations of this kind have previously been used by other workers;~\cite{christlee,Gaw}
It would be 
possible to make a compensating transformation on the states, but we prefer to leave the 
states untransformed and to extract 
\be
[{\sf H}]_0=\int\!d{\bf r}\left[ \ {\textstyle \frac{1}{2}}
{\cal P}^{b\,{\sf T}}_{j}({\bf r}){\cal P}^{b\,{\sf T}}_{j}({\bf r})
+  {\textstyle \frac{1}{4}} {\hat F}_{{\sf GI}\,ij}^a({\bf r}) {\hat F}_{{\sf GI}\,ij}^{a}({\bf r})+
{\psi^\dagger}({\bf r})\left(\beta m-i\alpha_{i}
\partial_{i}\right)\psi({\bf r})\right]\,,
\label{eq:h0h}
\ee
from Eq.~(\ref{eq:h1}) in order to obtain a non-interacting part of $({\hat { H}}_{\sf GI})_{\mbox{phys}}$ 
that consists of hermitian gauge-invariant fields and that can define interaction picture operators.
As we will show in Appendix~\ref{sec:appe0}, this process leads to the expression
\be
({\hat { H}}_{\sf GI})_{\mbox{phys}}=[{\sf H}]_0+[{\sf H}]_1+[{\sf H}]_2
\label{eq:Hnew}
\ee
where
\bea
&&\!\!\!\!\!\!\!\!\![{\sf H}]_1=\int\!d{\bf r}\left\{gf^{abc}\partial_iA^a_{{\sf GI}\,j}({\bf r})
A^b_{{\sf GI}\,i}({\bf r})A^c_{{\sf GI}\,j}({\bf r})+ 
{\textstyle \frac{1}{4}}g^2f^{abc}f^{ab^{\prime}c^{\prime}}
A^b_{{\sf GI}\,i}({\bf r})A^c_{{\sf GI}\,j}({\bf r})A^{b^{\prime}}_{{\sf GI}\,i}({\bf r})
A^{c^{\prime}}_{{\sf GI}\,j}({\bf r})\right.\nonumber \\
&&\!\!\!\!\!\!\!\!\!-\left.j^a_i({\bf r})A^a_{{\sf GI}\,i}({\bf{r}})\right\}-
{\ty \frac{1}{2}}\int\!d{\bf r}d{\bf x}d{\bf y}\left(j_0^b({\bf x})+
{\bar {\sf J}}_{0\,({\sf GI})}^{b\,{\sf T}}({\bf x})\right)
\stackrel{\longleftarrow}{{\cal D}^{\,ba}}({\bf x},{\bf r}){\partial^2}
{\cal D}^{\,ac}({\bf r},{\bf y})\left(j_0^c({\bf y})+
{\bar {\sf J}}_{0\,({\sf GI})}^{c\,{\sf T}}({\bf y})\right)\,
\label{eq:h1h}
\eea
and 
\bea
&&[{\sf H}]_2={\cal U}+{\cal V}+{\ty \frac{1}{2}}\int\!d{\bf r}d{\bf x}d{\bf y}\left\{
i{\sf k}^b_0({\bf x})\stackrel{\longleftarrow}{{\cal D}^{\,ba}}({\bf x},{\bf r}){\partial^2}
{\cal D}^{\,ac}({\bf r},{\bf y})\left(j_0^c({\bf y})+
{\bar {\sf J}}_{0\,({\sf GI})}^{c\,{\sf T}}({\bf y}))\right)-\right.\nonumber\\
&&\left.\left(j_0^b({\bf x})+
{\bar {\sf J}}_{0\,({\sf GI})}^{b\,{\sf T}}({\bf x})\right)
\stackrel{\longleftarrow}{{\cal D}^{\,ba}}({\bf x},{\bf r}){\partial^2}
{\cal D}^{\,ac}({\bf r},{\bf y})i{\sf k}^c_0({\bf y})+
{\sf k}^b_0({\bf x})\stackrel{\longleftarrow}{{\cal D}^{\,ba}}({\bf x},{\bf r}){\partial^2}
{\cal D}^{\,ac}({\bf r},{\bf y}){\sf k}^c_0({\bf y})\right\}\,,
\label{eq:h2h}
\eea
%where
%\be
%{\bar {\sf J}}_{0\,({\sf GI})}^{b\,{\sf T}}=gf^{buv}A^u_{{\sf GI}\,j}{\cal P}^{v\,{\sf T}}_{j}\,,
%\ee
where ${\cal U}$ and ${\cal V}$ as well as ${\sf k}^b_0({\bf x})$ and
${\bar {\sf J}}_{0\,({\sf GI})}^{b\,{\sf T}}$ are defined in Appendix~\ref{sec:appe0} 
in Eqs. (\ref{eq:U}), (\ref{eq:Ud}), (\ref{eq:JjPa3}), and (\ref{eq:Jbar}) respectively. 
$[{\sf H}]_0$, $[{\sf H}]_1$, and $[{\sf H}]_2$ are hermitian, and all consist entirely of 
gauge-invariant, hermitian, transverse gauge fields and gauge invariant quark fields, which all
obey ``standard'' commutation rules. Since ${\cal P}^{b\,{\sf T}}_{j}({\bf y})$
and $\Pi_{{\sf GI}\,j}^{b\,{\sf T\,\dagger}}({\bf y})$ have the same commutator with 
$A_{{\sf GI}\,i}^{a}({\bf{x}})$, Eq.~(\ref{eq:PATcomm}) also determines the commutation rule 
\bea
&&\left[A_{{\sf GI}\,j}^{b}({\bf{y}})\,,A_{{\sf GI}\,i}^{a}({\bf{x}})\right]
=\left[{\cal P}^{b\,{\sf T}}_{j}({\bf y})\,,{\cal P}^{a\,{\sf T}}_{i}({\bf x})\right]=0,\nonumber \\
&&\left[{\cal P}^{b\,{\sf T}}_{j}({\bf y})\,,A_{{\sf GI}\,i}^{a}({\bf{x}})\right]
=-i\delta_{ab}\left(\delta_{ij}-\frac{\partial_i\partial_j}{\partial^2}\right)
\delta({\bf x}-{\bf y})\,.
\label{eq:Phcomm}
\eea
The sum $[{\sf H}]_0+[{\sf H}]_1$ is identical in form to the Coulomb-gauge QCD Hamiltonian used
by Gribov~\cite{gribov,gribovb}, as well as by numerous other authors who have followed him in using
this Hamiltonian. $[{\sf H}]_2$ consists of additional terms that are required because the transverse gauge-invariant negative chromoelectric field
$\Pi_{{\sf GI}\,j}^{b\,{\sf T}}$ is not hermitian. The elimination of
$\Pi_{{\sf GI}\,j}^{b\,{\sf T}}$ and $\Pi_{{\sf GI}\,j}^{b\,{\sf T\,\dagger}}$ in 
favor of the hermitian ${\cal P}^{b\,{\sf T}}_{j}$ is essential for the
establishment of the isomorphism between $({\hat { H}}_{\sf GI})_{\mbox{phys}}$ and a Hamiltonian that can be 
used in a Fock space of perturbative states. 
We now proceed to the demonstration of this isomorphism.\s 

Since both $A_{{\sf GI}\,i}^{a}({\bf {r}})$ and ${\cal P}^{b\,{\sf T}}_{j}({\bf {r}})$ 
are hermitian and obey the commutation rule displayed in Eq.~(\ref{eq:Phcomm}), 
we can represent them as
\be
A_{{\sf GI}\,i}^c({\bf r})=\sum_{{\bf k},\,n}\frac{\epsilon_i{}^n({\bf k})}
{\sqrt{2k}}\left[{\alpha}^c_n({\bf k})e^{i{\bf
k\cdot r}} +{\alpha}_n^{c\,\dagger}({\bf k})e^{-i{\bf k\cdot r}}\right]
\label{eq:AGiT}
\ee  
and
\be
{\cal P}_{i}^{c\,{\sf T}}({\bf r})=-i\sum_{{\bf k},\,n}\epsilon_i{}^n({\bf
k})\sqrt{\frac{k}{2}}\left[{\alpha}^c_n({\bf k})e^{i{\bf
k\cdot r}} -{\alpha}_n^{c\,\dagger}({\bf k})e^{-i{\bf k\cdot r}}\right]
\label{eq:PGiT}
\ee
where $n$ is summed over two transverse helicity modes and
\be
\left[{\alpha}^a_n({\bf k})\,,{\alpha}_\ell^{b\,\dagger}({\bf q})\right]=
\delta_{n,\ell}\,\delta_{a,b}\,\delta_{{\bf k},{\bf q}}\;\;\mbox{and}
\;\;\left[{\alpha}^a_n({\bf k})\,,{\alpha}_\ell^{b}({\bf q})\right]=
\left[{\alpha}^{a\,\dagger}_n({\bf k})\,,{\alpha}_\ell^{b\,\dagger}({\bf q})\right]=0\,.
\label{eq:comalph}
\ee
Eqs.~(\ref{eq:AGiT}) and (\ref{eq:PGiT}) can be inverted, leading to 
\be
{\alpha}^c_n({\bf k})=\sqrt{\frac{k}{2}}\,\epsilon_i{}^n({\bf k}){\int}d{\bf r}\left(A_{{\sf GI}\,i}^c({\bf r})
+\frac{i}{k}{\cal P}_{i}^{c\,{\sf T}}({\bf r})\right)e^{-i{\bf k\cdot r}}
\label{eq:alph}
\ee
and
\be
{\alpha}_n^{c\,\dagger}({\bf k})=\sqrt{\frac{k}{2}}\,
\epsilon_i{}^n({\bf k}){\int}d{\bf r}\left(A_{{\sf GI}\,i}^c({\bf r})
-\frac{i}{k}{\cal P}_{i}^{c\,{\sf T}}({\bf r})\right)e^{i{\bf k\cdot r}}\,.
\label{eq:alphad}
\ee
Eqs.~(\ref{eq:alph}) and (\ref{eq:alphad}) show that ${\alpha}^c_n({\bf k})$ 
and ${\alpha}_n^{c\,\dagger}({\bf k})$
are gauge-invariant and commute with the Gauss's law operator ${\cal G}^a({\bf r})$.
Eqs.~(\ref{eq:ra}), (\ref{eq:rb}) and (\ref{eq:ppi}) demonstrate that any functional 
${\cal F}\left(A_{{\sf GI}\,i}^{a}\,,{\cal P}^{b\,{\sf T}}_{j}\right)$ 
will have the transformation property 
\be
{\cal F}\left(A_{{\sf GI}\,i}^{a}\,,{\cal P}^{b\,{\sf T}}_{j}\right)
{\cal J}^{-\frac{1}{2}}\Psi|\phi_{\ell}\rangle=
{\cal J}^{-\frac{1}{2}}\Psi\,{\cal F}\left(A_{i}^{a\,{\sf T}}\,,
{\Pi}^{b\,{\sf T}}_{j}\right)|\phi_{\ell}\rangle
\label{eq:Falph}
\ee
leading to 
\be
{\alpha}_n^{c}({\bf k}){\cal J}^{-\frac{1}{2}}\Psi|\phi_{\ell}\rangle=
{\cal J}^{-\frac{1}{2}}\Psi\,\left\{\sqrt{\frac{k}{2}}\,\epsilon_i{}^n({\bf k}){\int}d{\bf r}
\left(A_{i}^{c\,{\sf T}}({\bf r})
+\frac{i}{k}{\Pi}_{i}^{c\,{\sf T}}({\bf r})\right)e^{-i{\bf k\cdot r}}\right\}|\phi_{\ell}\rangle\,,
\label{eq:A1}
\ee
and 
\be
{\alpha}_n^{c\,\dagger}({\bf k}){\cal J}^{-\frac{1}{2}}\Psi|\phi_{\ell}\rangle=
{\cal J}^{-\frac{1}{2}}\Psi\,\left\{\sqrt{\frac{k}{2}}\,\epsilon_i{}^n({\bf k}){\int}d{\bf r}
\left(A_{i}^{c\,{\sf T}}({\bf r})
-\frac{i}{k}{\Pi}_{i}^{c\,{\sf T}}({\bf r})\right)e^{-i{\bf k\cdot r}}\right\}|\phi_{\ell}\rangle\,,
\label{eq:A1a}
\ee
so that the isomorphism established in Eq.~(\ref{eq:Falph}) between the gauge-invariant fields 
$A_{{\sf GI}\,i}^{a}$, ${\cal P}^{b\,{\sf T}}_{j}$ and the gauge-dependent Weyl-gauge fields
$A_{i}^{a\,{\sf T}}$, ${\Pi}^{b\,{\sf T}}_{j}$ respectively is transferred to a 
similar relation between the gauge-invariant creation and annihilation operators for 
transverse gluons, ${\alpha}_n^{c\,\dagger}({\bf k})$ and ${\alpha}_n^{c}({\bf k})$,
 and corresponding ``standard'' perturbative creation and annihilation 
operators ${a}_n^{c\,\dagger}({\bf k})$ and ${a}_n^{c}({\bf k})$.
We can proceed by using the standard representation for the transverse 
part of the Weyl-gauge fields, 
\be
A_{i}^{c\,{\sf T}}({\bf r})=\sum_{{\bf k},\,n}\frac{\epsilon_i{}^n({\bf k})}
{\sqrt{2k}}\left[{a}^c_n({\bf k})e^{i{\bf
k\cdot r}} +{a}_n^{c\,\dagger}({\bf k})e^{-i{\bf k\cdot r}}\right]
\label{eq:AWT}
\ee  
and
\be
\Pi_{i}^{c\,{\sf T}}({\bf r})=-i\sum_{{\bf k},\,n}\epsilon_i{}^n({\bf
k})\sqrt{\frac{k}{2}}\left[{a}^c_n({\bf k})e^{i{\bf
k\cdot r}} -{a}_n^{c\,\dagger}({\bf k})e^{-i{\bf k\cdot r}}\right]\,,
\label{eq:PWT}
\ee
which demonstrate that 
\bea
&&{\alpha}_n^{c\,\dagger}({\bf k}){\cal J}^{-\frac{1}{2}}\Psi|\phi_i\rangle=
{\cal J}^{-\frac{1}{2}}\Psi\,{a}_n^{c\,\dagger}({\bf k})|\phi_i\rangle\;\;\;\;\mbox{and}\nonumber \\
&&{\alpha}_n^{c}({\bf k}){\cal J}^{-\frac{1}{2}}\Psi|\phi_i\rangle=
{\cal J}^{-\frac{1}{2}}\Psi\,{a}_n^{c}({\bf k})|\phi_i\rangle\,.
\label{eq:A2}
\eea
Any ${\alpha}_n^{c}({\bf k})$ will annihilate the gauge-invariant vacuum state 
${\cal J}^{-\frac{1}{2}}\Psi\,\Xi|0\rangle$, because the transverse excitation operators 
${a}_n^{c}({\bf k})$ and ${a}_n^{c\,\dagger}({\bf k})$ trivially commute with $\Xi$.\s

At this point, we can establish an isomorphism between two Fock spaces: The ``standard'' Weyl-gauge Fock space 
consists of 
\be
|{\bf k}\rangle={a}_n^{c\,\dagger}({\bf k})|0\rangle
\label{eq:k1}
\ee
$$\cdots$$ 
\be|{\bf k}_1\cdots{\bf k}_i\cdots{\bf k}_N\rangle=K\left[{a}_{n_{1}}^{{c_{1}}\,\dagger}({\bf k}_1)\cdots
{a}_{n_{i}}^{{c_{i}}\,\dagger}({\bf k}_i)\cdots{a}_{n_{N}}^{{c_{N}}\,\dagger}({\bf k}_N)\right]|0\rangle\,;
\label{eq:kn}
\ee 
with $K$ the normalization constant
and the gauge-invariant states that implement the non-Abelian Gauss's law can be represented as
\be
|{\bf {\bar k}}\rangle={\ty \frac{1}{C}}{\alpha}_n^{c\,\dagger}({\bf k})
{\cal J}^{-\frac{1}{2}}\Psi\Xi|0\rangle
\ee
$$\cdots$$
\be
|{\bf {\bar k}}_1\cdots{\bf {\bar k}}_i\cdots{\bf {\bar k}}_N\rangle={\ty \frac{K}{C}}
\left[{\alpha}_{n_{1}}^{{c_{1}}\,\dagger}({\bf k}_1)\cdots
{\alpha}_{n_{i}}^{{c_{i}}\,\dagger}({\bf k}_i)\cdots{\alpha}_{n_{N}}^{{c_{N}}\,\dagger}({\bf k}_N)\right]
{\cal J}^{-\frac{1}{2}}\Psi\Xi|0\rangle
\ee
where $|0\rangle$ designates the perturbative vacuum annihilated by ${a}^c_n({\bf k})$ as well as by the
annihilation operators for quarks and antiquarks,  
$q_{{\bf p},s}$ and ${\bar q}_{{\bf p},s}$ respectively. The additional normalization constant
$C^{-1}$ must be introduced to compensate for the fact 
that $|C|^2=|{\cal J}^{-\frac{1}{2}}\Psi\Xi|0\rangle|^2=
\langle0|\Xi^\star\Psi^\star{\cal J}^{-1}\Psi\Xi|0\rangle$, which formally is a universal positive
constant, is not finite; and the state ${\cal J}^{-\frac{1}{2}}\Psi\Xi|0\rangle$  is not normalizable. 
However, once $C$ is introduced, the 
$|{\bf {\bar k}}_1\cdots{\bf {\bar k}}_i\cdots{\bf {\bar k}}_N\rangle$ 
states form a satisfactory Fock space that is gauge-invariant as well as 
isomorphic to the space of $|{\bf k}_1\cdots{\bf k}_i\cdots{\bf k}_N\rangle$ states. We can now use 
Eqs.~(\ref{eq:AGiT}), and (\ref{eq:PGiT}) to express $[{\sf H}]_0$ as
\be
[{\sf H}]_0=\sum_{{\bf k}, c}k{\alpha}_n^{c\,\dagger}({\bf k}){\alpha}_n^{c}({\bf k})+
\sum_{{\bf p},s}{\cal E}_{\bf p}\left(q_{{\bf p},s}^{\dagger}q_{{\bf p},s}+
{\bar q}_{{\bf p},s}^\dagger{\bar q}_{{\bf p},s}\right)\,
\label{eq:h0spect}
\ee
with the subscript $s$ labeling the color, flavor and herlicity of the quarks.
In this form, $[{\sf H}]_0$ can be seen to describe the energy of non-interacting gauge-invariant
transverse gluons of energy $k$ and quarks and anti-quarks respectively of energy ${\cal E}_{\bf p}=\sqrt{m^2+|{\bf p}|^2}$.
%%%%%%%%%%%%%%%%%%%%%%%%%%
We can also define another Hamiltonian, 
${\cal H}={\cal H}_0+{\cal H}_1+{\cal H}_2$, in which each component part is identical in form to 
$[{\sf H}]_0+[{\sf H}]_1+[{\sf H}]_2$ respectively, but with the substitutions 
$$
{\cal P}^{b\,{\sf T}}_{j}\ra\Pi^{b\,{\sf T}}_{j}\;\;\;\mbox{and}\;\;\;
A^a_{{\sf GI}\,i}{\ra} A^{a\,{\sf T}}_{i}
$$
everywhere --- including the replacement of $A^a_{{\sf GI}\,i}$ by $A^{a\,{\sf T}}_{i}$ 
in the inverse 
Faddeev-Popov operator ${\cal D}^{\,ab}({\bf x},{\bf y})$ ---
so that ${\cal H}$ is characteristic of the
Coulomb gauge, but nevertheless is a  functional of transverse Weyl-gauge unconstrained 
fields. For example, ${\cal H}_0$
is
\be
{\cal H}_0=\int d{\bf r} \left[\ {\textstyle \frac{1}{2}
\Pi^{a\,{\sf T}}_{i}({\bf r})\Pi^{a\,{\sf T}}_{i}({\bf r})
+ \frac{1}{4}} {\hat F}_{ij}^{a}({\bf r}) {\hat F}_{ij}^{a}({\bf r})+
{\psi^\dagger}({\bf r})
\left(\,\beta m-i\alpha_{i}
\partial_{i}\right)
\psi({\bf r})\right]\,
\label{eq:HQCDC}
\ee
where ${\hat F}_{ij}^a=\partial_jA^a_{i}-\partial_iA^a_{j}$. Using Eqs.~(\ref{eq:AWT}), and 
(\ref{eq:PWT}), we can express 
${\cal H}_0$ in the form
\be
{\cal H}_0=\sum_{{\bf k}, c}k{a}_n^{c\,\dagger}({\bf k}){a}_n^{c}({\bf k})+
\sum_{{\bf p},s}{\cal E}_{\bf p}\left(q_{{\bf p},s}^{\dagger}q_{{\bf p},s}+
{\bar q}_{{\bf p},s}^\dagger{\bar q}_{{\bf p},s}\right)\,.
\label{eq:h0wspec}
\ee 
We can then use Eq.~(\ref{eq:Falph}) to establish that 
\be
[{\sf H}]_0
{\cal J}^{-\frac{1}{2}}\Psi\,\Xi|n\rangle=
{\cal J}^{-\frac{1}{2}}\Psi\,\Xi\,{\cal H}_0|n\rangle
\label{eq:hotran}
\ee
as well as 
\be
[{\sf H}]_1
{\cal J}^{-\frac{1}{2}}\Psi\,\Xi|n\rangle=
{\cal J}^{-\frac{1}{2}}\Psi\,\Xi\,{\cal H}_1|n\rangle
\label{eq:hotran1}
\ee
and 
\be
[{\sf H}]_2
{\cal J}^{-\frac{1}{2}}\Psi\,\Xi|n\rangle=
{\cal J}^{-\frac{1}{2}}\Psi\,\Xi\,{\cal H}_2|n\rangle
\label{eq:hotran2}
\ee
The state vector 
$|n\rangle$ represents one of the $|{\bf k}_1\cdots{\bf k}_i\cdots{\bf k}_N\rangle$, the 
``standard'' perturbative eigenstates of ${\cal H}_0$.\s

We can use the relations between Weyl-gauge and gauge-invariant states we have established in the preceding discussion
to extend the isomorphism we have demonstrated to include scattering transition amplitudes. 
For this purpose, we define 
\be
 {\cal H}_{\rm int}={\cal H}_1+{\cal H}_2\;\;\;\mbox{and}\;\;\;[{\sf H}]_{\rm int}=[{\sf
H}]_1+[{\sf H}]_2\,.
\label{eq:hinttran}
\ee

The transition
amplitude between gauge-invariant states is given by 
\be
\overline{{\sf T}}_{f,i}=\frac{1}{C^2}{\langle}f|\Xi^\star\Psi^\star{\cal J}^{-\frac{1}{2}}
\left\{[{\sf H}]_{\rm int}+[{\sf H}]_{\rm int}
\frac{1}{\left(E_i-({\hat { H}}_{\sf GI})_{\mbox{phys}}+i\epsilon\right)}
[{\sf H}]_{\rm int}\right\}{\cal J}^{-\frac{1}{2}}\Psi\Xi|i\rangle\,,
\label{eq:hyb1}
\ee
where $|i\rangle$ and $|f\rangle$ each designate one of the $|n\rangle$ states; $|i\rangle$ represents an incident
and $|f\rangle$ a final state in a scattering process. With the results of the preceding discussion, we can express this as
\bea
\overline{{\sf T}}_{f,i}\!\!\!\!\!\!&&=\frac{1}{C^2}{\langle}f|\Xi^\star
\Psi^\star{\cal J}^{-1}\Psi\Xi|n\rangle{\langle}n|
\left\{ {\cal H}_{\rm int}+ {\cal H}_{\rm int}
\frac{1}{\left(E_i-{\cal H}_0- {\cal H}_{\rm int}+i\epsilon\right)}
{\cal H}_{\rm int}\right\}|i\rangle\nonumber \\
&&=\frac{1}{C^2}{\langle}0|\Xi^\star\Psi^\star{\cal J}^{-1}\Psi\Xi|0\rangle{\langle}f|
\left\{ {\cal H}_{\rm int}+ {\cal H}_{\rm int}
\frac{1}{\left(E_i-{\cal H}_0- {\cal H}_{\rm int}+i\epsilon\right)}
{\cal H}_{\rm int}\right\}|i\rangle
\label{eq:hyb1}
\eea
where we sum over the complete set of perturbative states $|n\rangle{\langle}n|$. The second line of
Eq.~(\ref{eq:hyb1}) follows from 
\bea
&&\frac{1}{C^2}{\langle}f|\Xi^\star
\Psi^\star{\cal J}^{-1}\Psi\Xi|n\rangle=\frac{1}{C^2}{\langle}0|a_{f}({\bf k}_f)\Xi^\star
\Psi^\star{\cal J}^{-1}\Psi{\Xi}a^\dagger_n({\bf k}_n)|0\rangle=\frac{1}{C^2}{\langle}0|\Xi^\star
\Psi^\star{\cal J}^{-\frac{1}{2}}\alpha_{f}({\bf k}_f)\alpha^\dagger_n({\bf k}_n)
{\cal J}^{-\frac{1}{2}}\Psi{\Xi}|0\rangle=\nonumber\\
&&
\delta_{f,n}{\delta}({\bf k}_f-{\bf k}_n)\frac{1}{C^2}{\langle}0|\Xi^\star
\Psi^\star{\cal J}^{-1}\Psi\Xi|0\rangle-\frac{1}{C^2}{\langle}0|\Xi^\star
\Psi^\star{\cal J}^{-\frac{1}{2}}\alpha^\dagger_n({\bf k}_n)\alpha_{f}({\bf k}_f)
{\cal J}^{-\frac{1}{2}}\Psi{\Xi}|0\rangle
\label{eq:hyb3}
\eea
and the observation that the last term on the second line of Eq.~(\ref{eq:hyb3}) vanishes trivially.
With the isomorphism of the states 
$|{\bf k}_1\cdots{\bf k}_i\cdots{\bf k}_N\rangle$ and 
$|{\bf {\bar k_1}}\cdots{\bf {\bar k_i}}\cdots{\bf {\bar k_N}}\rangle$ that we have established, 
\be
\overline{{\sf T}}_{f,i}=T_{f,i}
\ee
where 
\be
T_{f,i}={\langle}f|
\left\{ {\cal H}_{\rm int}+ {\cal H}_{\rm int}
\frac{1}{\left(E_i-{\cal H}_0- {\cal H}_{\rm int}+i\epsilon\right)}
 {\cal H}_{\rm int}\right\}|i\rangle
\label{eq:Treg}
\ee
is a transition amplitude 
that can be evaluated with Feynman graphs and rules, because it is based on ``standard'' 
perturbative states that are not required to implement Gauss's law and need not be gauge-invariant.\s

In the remainder of this section, we will discuss the relation of our formulation of 
the scattering transition amplitude to approaches to this problem 
in Coulomb-gauge formulations of QCD. 
As was pointed out in section IIB, the effective Hamiltonian $(\hat H_{\sf GI})_{\rm phys}$ 
described in Eq.~(\ref{eq:Heff}) is identical to one obtained by Christ and Lee, \cite{christlee}
who treated gauge fields as coordinates and 
applied the apparatus of transformations from Cartesian to curvilinear coordinates 
to the problem of formulating Coulomb-gauge QCD. Here, we will show that $\hat H_{\sf GI}$ ---
the precursor of $(\hat H_{\sf GI})_{\rm phys}\,$, described in Eq.~(\ref{eq:HQCDN}) --- is identical in form 
to the Hamiltonian given in Eq.~(6.15) in Ref.~\cite{christlee}, which leads to the Coulomb-gauge
perturbative rules formulated by Christ and Lee. For this purpose, $\hat H_{\sf GI}$ will be 
expressed in terms of ${\cal P}_j^{b\,{\sf T}}$ and $A_{{\sf GI}\,i}^a$, and then Weyl-ordered.   
The equivalence of Christ and Lee's results with 
Schwinger's~\cite{schwingerb} was already confirmed in Ref.~\cite{christlee}.\s

Eq.~(\ref{eq:hermpb}) demonstrates that
the functional dependence of 
$\hat H_{\sf GI}$ on ${\cal P}_j^{b\,{\sf T}}$ and $A_{{\sf GI}\,i}^a$ 
is the same as the functional dependence of 
\be
\bar H\equiv {\cal J}^{1\over 2}\hat H_{\sf GI}{\cal J}^{-{1\over 2}}
\label{eq:Hbar}
\ee
on $\Pi_{j}^{b\,{\sf T}}$ and $A_{{\sf GI}\,i}^a$. 
$\bar H$ was used by Christ and Lee to generate the
path  integral representation of the Coulomb gauge,~\cite{christlee} and they showed that
\be
\bar H=\hat H_{\sf GI}^W+{\cal V}_1+{\cal V}_2
\label{eq:HbarWeyl}
\ee 
where the superscript $W$ designates Weyl-ordering with respect to $\Pi_{{\sf GI}\,j}^{b\,{\sf T}}$ and
$A_{{\sf GI}\,i}^a$. The additional terms ${\cal V}_1$ and ${\cal V}_2$ are given by 
\be
{\cal V}_1={1\over 8}g^2f^{lbc}f^{lad}\int d{\bf r}{\partial_j}{\cal D}^{ab}({\bf r},{\bf r})
{\partial_j}{\cal D}^{cd}({\bf r},{\bf r})
\label{eq:V1}
\ee
and
\bea
{\cal V}_2=&&{1\over 8}g^2f^{lna}f^{kbm}\int d{\bf x}d{\bf y}d{\bf z}
\left(\delta_{kn}\delta_{ji}\delta({\bf y}-{\bf x})+D_j{\cal D}^{kn}({\bf y},{\bf x})
\stackrel{\leftarrow}{\partial}_i\right)\times \nonumber \\
&&{\cal D}^{ac}({\bf x},{\bf z})\partial^2
{\cal D}^{cb}({\bf z},{\bf y}))
\left(\delta_{lm}\delta_{ij}\delta({\bf x}-{\bf y})+D_i{\cal D}^{lm}({\bf x},{\bf y})
\stackrel{\leftarrow}{\partial}_j\right)
\label{eq:V2}
\eea
where the partial derivative $\partial_j$ to the left of ${\cal D}^{ab}({\bf r},{\bf r})$ 
acts only on its first argument. When a partial derivative with a left arrow on top appears to the
right of ${\cal D}^{ab}$ with two identical arguments, it acts only on its second argument. The case of of 
two identical arguments of ${\cal D}$ is understood as  
$\partial_j{\cal D}^{ab}({\bf x},{\bf x})\equiv\lim_{{\bf y}\to{\bf x}}
{\partial\over\partial x_j}{\cal D}^{ab}({\bf x},{\bf y})$ and 
${\cal D}^{ab}({\bf x},{\bf x})\stackrel{\leftarrow}{\partial}_j\equiv\lim_{{\bf y}\to{\bf x}}
{\partial\over\partial x_j}{\cal D}^{ab}({\bf y},{\bf x})$, where the limit is taken {\em after}
the  partial derivative has been evaluated. 
This convention will
be  followed consistently in the following discussions. Since the commutator of 
${\cal P}_j^{b\,{\sf T}}$ and $A_{{\sf GI}\,i}^a$ is identical to that of
$\Pi_{{\sf GI}\,j}^{b\,{\sf T}}$ and $A_{{\sf GI}\,i}^a\,$, an equation parallel to Eq.~(\ref{eq:HbarWeyl}), 
\be
\hat H_{\sf GI}=\hat H_{\sf GI}^W+{\cal V}_1+{\cal V}_2
\label{eq:HWeyl}
\ee 
will be proven below. The superscript $W$ designates Weyl-ordering, but in this case  with respect to
${\cal P}_j^{b\,{\sf T}}$ and $A_{{\sf GI}\,i}^a\,$. The parallel structure refers to the fact that, 
as was pointed out above,
$\hat H_{\sf GI}$ has the same functional dependence on ${\cal P}_j^{b\,{\sf T}}$ and $A_{{\sf
GI}\,i}^a$ as
$\bar H$ has on $\Pi_{{\sf GI}\,j}^{b\,{\sf T}}$ and $A_{{\sf GI}\,i}^a$. 
Since the fermion variables commute with ${\cal P}_j^{b\,{\sf T}}$ and $A_{{\sf GI}\,i}^a$, we may drop them for
the  proof of Eq.~(\ref{eq:HWeyl}); we will also drop $H_{\cal G}$, since it makes no contributions 
 in the space
of gauge-invariant states. It follows from Eq.~(\ref{eq:Pirepc}) that for a physical state
$|\Psi\rangle$,
\be
\Pi_{{\sf GI}\,j}^b({\bf r})|\Psi\rangle=-{\cal E}_j^b({\bf r})|\Psi\rangle
\label{eq:colorE}
\ee
with 
\be
{\cal E}_j^b({\bf r})={\cal J}^{1\over 2}\Big[-{\cal P}_j^{b\,{\sf T}}({\bf r})
+\int d{\bf x}\partial_j{\cal D}^{bq}({\bf r},{\bf x})D_i({\bf x})
{\cal P}_i^{q\,{\sf T}}({\bf x})\Big]{\cal J}^{-{1\over 2}}\,.
\label{eq:colorEq}
\ee
In terms of the Weyl-ordered chromoelectric field operator of Schwinger,~\cite{schwingera}
\be
E_j^b({\bf r})=-{\cal P}_j^{b\,{\sf T}}({\bf r})
+{\ty {1\over 2}}\int d{\bf x}\Big[\partial_j{\cal D}^{bq}({\bf r},{\bf x})
D_i({\bf x}){\cal P}_i^{q\,{\sf T}}({\bf x})
+D_i({\bf x}){\cal P}_i^{q\,{\sf T}}({\bf x})\partial_j{\cal D}^{bq}({\bf r},{\bf x})\Big]\,,
\label{eq:colorEw}
\ee
we have
\bea
&&{\cal E}_j^b({\bf r})=E_j^b({\bf r})+\Delta_j^b({\bf r})\;\;\;\mbox{and}\nonumber \\
&&{\cal E}_j^b({\bf r})^\dagger=E_j^b({\bf r})-\Delta_j^b({\bf r}).
\label{eq:Delta}
\eea
The Hamiltonian $\hat H_{\sf GI}$, in the absence of the fermion field and without 
$H_{\cal G}$, can be written as
\be
\hat H_{\sf GI}=K+{1\over 4}\int d{\bf r}F_{{\sf GI}ij}^a({\bf r})F_{{\sf GI}ij}^a({\bf r})
\label{eq:rearrange}
\ee
where the kinetic energy
\bea
K&&={1\over 2}\int d{\bf r}{\cal E}_j^b({\bf r})^\dagger{\cal E}_j^b({\bf r})\nonumber \\
&&={1\over 2}\int d{\bf r}[E_j^b({\bf r})E_j^b({\bf r})+v({\bf r})]
\label{eq:kinetic}
\eea
with
\be
v({\bf r})=-\Delta_j^b({\bf r})\Delta_j^b({\bf r})+\Big[E_j^b({\bf r}),\Delta_j^b({\bf r})\Big].
\label{eq:integrand}
\ee
To evaluate $\Delta_j^b({\bf r})$, we observe that 
\be
{\cal E}_j^b({\bf r})=-{\cal P}_j^{b\,{\sf T}}({\bf r})+{\ty {1\over2}}\left[{\cal P}_j^{b\,{\sf T}}({\bf
r})\,,\ln({\cal J}) \right] +\int d{\bf x}\partial_j{\cal D}^{bq}({\bf r},{\bf x})D_i({\bf x})
{\cal P}_i^{q\,{\sf T}}({\bf x})-{\ty {1\over2}}\int d{\bf x}\partial_j{\cal D}^{bq}({\bf r},{\bf x})D_i\,
\left[{\cal P}_i^{q\,{\sf T}}({\bf x})\,,\ln({\cal J}) \right]
\label{eq:calE1}
\ee
and
\be
{\cal E}_j^b({\bf r})^\dagger=-{\cal P}_j^{b\,{\sf T}}({\bf r})-{\ty {1\over2}}\left[{\cal P}_j^{b\,{\sf T}}({\bf
r})\,,\ln({\cal J}) \right] +\int d{\bf x}D_i({\bf x}){\cal P}_i^{q\,{\sf T}}({\bf x})\partial_j{\cal D}^{bq}({\bf r},
{\bf x})+{\ty {1\over2}}\int d{\bf x}\partial_j{\cal D}^{bq}({\bf r},{\bf x})D_i\,
\left[{\cal P}_i^{q\,{\sf T}}({\bf x})\,,\ln({\cal J}) \right]
\label{eq:calE2}
\ee 
so that 
\be
{\ty {1\over2}}\left({\cal E}_j^b({\bf r})+{\cal E}_j^b({\bf r})^\dagger\right)=E_j^b({\bf r})
\label{eq:calE3}
\ee
and, therefore, that 
\be
{\ty {1\over2}}\left({\cal E}_j^b({\bf r})-{\cal E}_j^b({\bf r})^\dagger\right)=\Delta_j^b({\bf r})\,.
\label{eq:calE4}
\ee
With Eqs.~(\ref{eq:colorE}) and (\ref{eq:Tcomm}), this leads to 
\be
\Delta_j^b({\bf r})=-\frac{i}{2}gf^{bch}{\frac{\partial}{{\partial}r_j}}{\cal D}^{\,ch}({\bf r},{\bf r})
\ee
In Appendix~\ref{sec:Appf}, we shall prove that
\be
{1\over 2}\int d{\bf r}v({\bf r})={\cal V}_1.
\label{eq:final1}
\ee
In the form given in Eq.~(\ref{eq:rearrange}) with $K$ as described in Eq.~(\ref{eq:kinetic}), 
the effective Hamiltonian $(\hat H_{\sf GI})$ is identified with that of
Schwinger.~\cite{schwingerb}  The next step towards 
the proof of (\ref{eq:HWeyl}) follows from the operator identity given in Ref.~\cite{christlee}
\be
{1\over 2}\int d{\bf r}E_j^b({\bf r})E_j^b({\bf r})=
{1\over 2}\int d{\bf r}\Big[E_j^b({\bf r})E_j^b({\bf r})\Big]^W
-{1\over 8}\int d{\bf x}d{\bf y}d{\bf z}\Big[{\cal P}_i^{a\,{\sf T}}({\bf x}),D_k{\cal D}^{bc}({\bf x},{\bf z})
\stackrel{\longleftarrow}{\partial_j}\Big]\Big[{\cal P}_k^{b\,{\sf T}}({\bf y}),D_i{\cal D}^{ac}({\bf y},{\bf z})
\stackrel{\longleftarrow}{\partial_j}\Big].
\label{eq:second}
\ee
Using the commutation relation (\ref{eq:Uc}), we can show that the second term on the right hand side of 
Eq.~(\ref{eq:second}) is the same as ${\cal V}_2$ (the same proof is also given in Ref.~\cite{christlee}) 
and Eq.~(\ref{eq:HWeyl}) is established.

\section{Discussion}
\label{sec:discuss}
In this work, we have used earlier results~\cite{CBH2,BCH3,HGrib} to express the Weyl-gauge
Hamiltonian entirely in terms of operator-valued fields that are gauge-invariant as well as 
 path-independent.  These gauge-invariant fields have many features in common with Coulomb-gauge 
fields: Their commutation rules agree with those given by  Schwinger 
in his Coulomb-gauge formulation of QCD,~\cite{schwingera,schwingerb}
except for differences in operator order; these differences can be ascribed to the fact that Schwinger
imposed Weyl  order in his work while we do not make any {\em ad hoc} changes in operator order. 
The gauge-invariant gauge field is transverse and hermitian; but the gauge-invariant
chromoelectric field is neither transverse nor hermitian. Even the transverse part of the gauge-invariant
chromoelectric field is not hermitian. That fact is important for relating the Hamiltonian we obtained in 
Eq.~(\ref{eq:Hnew}) with those given by Gribov,~\cite{gribov} Schwinger,~\cite{schwingerb} and Christ and Lee.~\cite{christlee}\s

The relation between the Coulomb-gauge Hamiltonian for QCD and the 
Weyl-gauge Hamiltonian expressed in terms
of gauge-invariant fields closely parallels the relation between the two corresponding 
QED Hamiltonians. The Weyl-gauge Hamiltonian for QCD is 
represented entirely in terms of gauge-invariant
fields in Eqs.~(\ref{eq:HQCDN}) and  (\ref{eq:HamGauss}).  
When formulated in terms of gauge-invariant fields, QCD  
must be embedded in a space of gauge-invariant states that obey the non-Abelian  Gauss's law. 
Within such a space of gauge-invariant states, further transformation of 
the QCD Hamiltonian we have constructed can be effected. Thus transformed, the
Hamiltonian consists of two parts. One part, $({\hat { H}}_{\sf GI})_{\mbox{phys}}$ ---
displayed in Eq.~(\ref{eq:Heff}) --- is identical to the Coulomb-gauge Hamiltonian. It is a 
functional of transverse gauge-invariant chromoelectric fields,
gauge-invariant gauge fields (which are inherently transverse), as well as gauge-invariant quark fields.
The other part, $H_{\cal G}$ --- displayed in Eq.~(\ref{eq:HamGauss}) --- makes only vanishing contributions
to matrix elements within the space of gauge-invariant states that is required for the Hamiltonian  
to act consistently as the time-evolution operator. $H_{\cal G}$ does affect the 
field equations and ``remembers'' that the formulation 
is for the Weyl, and not the Coulomb gauge. This situation is precisely the same as in QED,
in which the Weyl-gauge Hamiltonian, expressed in terms of the gauge-invariant field (in that case, 
simply the transverse part of the gauge field), is the sum of two terms, given in 
Eqs.~(\ref{eq:Hqedtc}) and (\ref{eq:Hg}); the former is the Coulomb-gauge Hamiltonian, and the latter
makes only vanishing contributions to matrix elements within the space of gauge-invariant states, but 
is necessary for reproducing the Euler-Lagrange equations for Weyl-gauge QED.  \s

In spite of the similarity between QCD and QED in the relation between the Weyl and Coulomb
gauges summarized in the preceding paragraph, there is an important difference between the
gauge-invariant states for the two theories:
Gauge-invariant and perturbative states in QED are unitarily equivalent; and 
in a Hamiltonian formulation, this unitary equivalence permits us to use perturbative 
states in evaluating scattering amplitudes in QED in algebraic and covariant gauges 
without compromising the implementation of Gauss's law.~\cite{khqedtemp,{khelqed}} 
But there can be no unitary equivalence between gauge-invariant states and 
perturbative states in QCD. And the gauge-invariant states in QCD are complicated, not
normalizable, and very cumbersome to use. In order to make effective use of the 
Weyl-gauge QCD Hamiltonian represented in terms of gauge-invariant fields, some relation 
is required that allows us to circumvent the absence of the unitary equivalence between 
gauge-invariant and perturbative states that afflicts non-Abelian gauge theories. 
In Section~\ref{sec:pert} we establish such a relation 
in the form of an isomorphism that enables us to consistently carry out calculations 
in QCD with an equivalent Hamiltonian that is a functional of 
the original gauge-dependent Weyl-gauge fields 
and that is used with standard perturbative states. 
In the case of QCD, this isomorphism has been demonstrated for the Weyl gauge only. An extension to 
a somewhat larger class of algebraic gauges defined by $A_0+{\gamma}A_3=0$ with $\gamma{\ge}0$ 
should not be difficult;~\cite{khgax}
but, in contrast to QED,  there is no indication that further extensions --- to covariant gauges, for example --- are
possible.  Finally, in Section~\ref{sec:pert}, we show that the effective Hamiltonian $(\hat H_{\sf GI})_{\rm phys}$
--- and therefore also ${\cal H}={\cal H}_0+{\cal H}_1+{\cal H}_2$ --- can be expressed in appropriately 
Weyl-ordered forms and shown to be equivalent to results obtained by Schwinger~\cite{schwingerb} and 
by Christ and Lee~\cite{christlee}. The Hamiltonian used by Gribov in Ref.~\cite{gribov} is equivalent to 
only ${\cal H}={\cal H}_0+{\cal H}_1$. ${\cal H}_2$ does not appear in that work, because the nonhermiticity of the
transverse chromoelectric field was not taken into account.

\section{Acknowledgments}
One of us (KH) thanks Prof. Daniel Zwanziger
for a helpful conversation and a written communication, Dr. Michael Creutz for a helpful written
communication, and  Profs. Carl Bender and
Gerald Dunne for helpful conversations.
The research of KH was supported by the Department of Energy under Grant
No.~DE-FG02-92ER40716.00 and that of HCR was supported by the Department of Energy under Grant
No.~DE-FG02-91ER40651.TASK B.

\appendix
\section{}
\label{subsec:FPinv}
In this section, we will prove Eqs.~(\ref{eq:FPi}) and (\ref{eq:FPd}). 
We use  Eqs. (\ref{eq:Pihermg})-(\ref{eq:Da}) and expand the product 
${\cal D}^{{a}h}({\bf y},{\bf x})\stackrel{\Longleftarrow}{\partial{\cdot}D^{hb}}_{({\bf x})}$
in Eq. (\ref{eq:FPi}) as a series in powers of $g$, and observe that ${\sf O}(n)$ terms originate from
${\cal D}^{{a}h}_{(n)}({\bf y},{\bf x})\left({\delta_{hb}}\stackrel{\longleftarrow}{\partial^2}_{({\bf x})}\right)$ 
and from ${\cal D}^{{a}h}_{(n-1)}({\bf y},{\bf x})\left(-gf^{hqb}\stackrel{\leftarrow}{\partial_i}A^q_{{\sf GI}\,i}\right)$.
For example, the first part of the $n=0$ term of Eq. (\ref{eq:FPi}) originates from the $\delta_{hb}
\stackrel{\longleftarrow}{\partial^2}$ part of $\stackrel{\Longleftarrow}{\partial{\cdot}D^{hb}}({\bf x})$ and is 
\be
\frac{-\delta_{ah}}{{4{\pi}|{\bf y}-{\bf x}|}}\left(\delta_{hb}
\stackrel{\longleftarrow}{\partial^2}_{({\bf x})}\right)=
{\partial^2}_{({\bf x})}\frac{-\delta_{ab}}{4\pi|{\bf x}-{\bf y}|}={\delta_{ab}}\delta({\bf x}-{\bf y})
\ee
and the second part of the $n=0$ term of Eq. (\ref{eq:FPi}), which stems from the 
$-gf^{hqb}\stackrel{\leftarrow}{\partial_i}A^q_{{\sf GI}\,i}$ in 
$\stackrel{\Longleftarrow}{\partial{\cdot}D^{hb}}({\bf x})$, is 
\be
\frac{-\delta_{ah}}{4\pi|{\bf y}-{\bf x}|}\left(-gf^{hqb}\stackrel{\leftarrow}{\partial_i}A^q_{{\sf GI}\,i}({\bf x})\right)
=-gf^{aqb}{\partial_{i\,({\bf y})}}\frac{1}{4\pi|{\bf y}-{\bf x}|}A^q_{{\sf GI}\,i}({\bf x})\,.
\ee
The first part of the $n=1$ term of Eq. (\ref{eq:FPi})
\be
\left\{gf^{qah}\int\frac{d{\bf z}}{{4{\pi}|{\bf y}-{\bf z}|}}
A_{{\sf GI}\;j}^{q}({\bf z})\frac{\partial}{\partial z}_j\left(\frac{1}{{4{\pi}|{\bf z}-{\bf x}|}}\right)\right\}
\delta_{hb}\stackrel{\longleftarrow}{\partial^2}_{({\bf x})}=gf^{aqb}{\partial_{i\,({\bf y})}}
\frac{1}{4\pi|{\bf y}-{\bf x}|}A^q_{{\sf GI}\,i}({\bf x})
\ee
 exactly cancels the second part of the $n=0$ term, and this 
pattern of cancellation can easily be seen to hold in general --- the first part of the $n+1$ term cancelling the 
second part of the $n$-th term. For the general term, 
\bea
&&{\cal D}^{{a}h}_{(n)}({\bf y},{\bf x})\delta_{hb}\stackrel{\longleftarrow}{{\partial^2}({\bf x})}=-f^{\alpha_1as_1}
f^{s_1\alpha_2s_2}{\cdots}f^{s_{n-1}\alpha_nb}g^n
{\int}\frac{d{\bf z}({\scriptstyle 1})}{4{\pi}|{\bf y}-
{\bf z}({\scriptstyle 1})|}A_{{\sf GI}\,l_1}^{{\alpha}_1}
({\bf z}({\scriptstyle 1}))\frac{\partial}{{\partial}z({\scriptstyle 1})_{l_1}}{\times}\nonumber \\
&&\!\!\!\!\!\!\!\!\!{\int}\frac{d{\bf z}({\scriptstyle 2})}
{4{\pi}|{\bf z}({\scriptstyle 1})-{\bf z}({\scriptstyle 2})|}
{\cdots}A_{{\sf GI}\;l}^{{\alpha}_{(n-1)}}
({\bf z}({\scriptstyle (n-1)}))\frac{\partial}{{\partial}z({\scriptstyle n-1})_l}
\frac{\partial}{{\partial}z({\scriptstyle (n-1)})_k}\;
\frac{1}{4{\pi}|{\bf z}({\scriptstyle n-1})-
{\bf x}|}A_{{\sf GI}\;k}^{{\alpha}_n}({\bf x})
\label{eq:calDna}
\eea 
and $\{{\cal D}^{{a}h}_{(n-1)}({\bf y},{\bf x})\}\{-gf^{hqb}\stackrel{\leftarrow}{\partial}_iA_{{\sf GI}\,i}^{b}({\bf x})\}$
is 
\bea
&&-{\cal D}^{{a}h}_{(n-1)}({\bf y},{\bf x})gf^{hqb}\stackrel{\longleftarrow}{\partial}_iA_{{\sf GI}\,i}^{b}({\bf x})
=f^{\alpha_1as_1}f^{s_1\alpha_2s_2}{\cdots}f^{s_{n-2}\alpha_{n-1}h}f^{hqb}g^n
{\int}\frac{d{\bf z}({\scriptstyle 1})}{4{\pi}|{\bf y}-
{\bf z}({\scriptstyle 1})|}A_{{\sf GI}\,l_1}^{{\alpha}_1}
({\bf z}({\scriptstyle 1}))\frac{\partial}{{\partial}z({\scriptstyle 1})_{l_1}}{\times}\nonumber \\
&&\!\!\!\!\!\!\!\!\!{\int}\frac{d{\bf z}({\scriptstyle 2})}
{4{\pi}|{\bf z}({\scriptstyle 1})-{\bf z}({\scriptstyle 2})|}
A_{{\sf GI}\,l_2}^{{\alpha}_2}
({\bf z}({\scriptstyle 2}))\frac{\partial}{{\partial}z{\scriptstyle (2)}_{l_2}}\;
{\cdots}A_{{\sf GI}\,l}^{{\alpha}_n}
({\bf z}{\scriptstyle (n-1)})\frac{\partial}{{\partial}z{\scriptstyle (n-1)}_{l}}
\frac{\partial}{{\partial}z{\scriptstyle (n-1)}_k}\frac{1}{{4{\pi}|{\bf z}{\scriptstyle (n-1)}-{\bf x}|}}
A_{{\sf GI}\;k}^{q}({\bf x})
\label{eq:calDn1b}
\eea 
so that, relabeling dummy indices $h{\rightarrow}s_{n-1}$ and $q\rightarrow\alpha_n$, we obtain  
$${\cal D}^{{a}h}_{(n)}({\bf y},{\bf x})\stackrel{\longleftarrow}{{\partial^2}_{({\bf x})}}
-{\cal D}^{{a}h}_{(n-1)}({\bf y},{\bf x})gf^{hqb}\stackrel{\leftarrow}{\partial}_iA_{{\sf GI}\,i}^{b}({\bf x})=0$$
and a consistent pattern of cancellations is established with ${\delta_{ab}}\delta({\bf y}-{\bf x})$
remaining as the only surviving term in  ${\cal D}^{{a}h}({\bf y},{\bf x})
\stackrel{\Longleftarrow}{\partial{\cdot}D^{hb}}({\bf x})$. A similar argument can be used to demonstrate Eq. (\ref{eq:FPd}).

\section{}
\label{sec:Appb}
In this section, we shall prove the identity given in Eq.~(\ref{eq:reneq2}).
By definition --- Eq.~(\ref{eq:Pigi2}) --- we have
\be
{\cal J}^{-1}\Pi_{{\sf GI}j}^b({\bf x}){\cal J}=\Pi_{{\sf GI}j}^b({\bf x})
+I_j^b({\bf x})\label{eq:B1}
\ee 
with 
\be
I_j^b({\bf x})={\cal J}^{-1}\left[\Pi_{{\sf GI}j}^b({\bf x})\,,{\cal J}\right]=
R_{bc}({\bf x}){\cal J}^{-1}\left[\Pi_{j}^c({\bf x})\,,{\cal J}\right]
\ee
so that
\bea
&&I_j^b({\bf x})=
-iR_{bc}({\bf x}){\delta \ln{\cal J}\over\delta A_j^c({\bf x})}\nonumber \\
&&=-iR_{bc}({\bf x})
\int d{\bf y}\int d{\bf z}{\cal D}^{mn}({\bf y},{\bf z})
{\delta\over\delta A_j^c({\bf x})} 
{\partial}\cdot{ D}^{nm}({\bf z})\delta({\bf z}-{\bf y})\nonumber \\
&&=igf^{lnm}R_{bc}({\bf x})
\int d{\bf y}\int d{\bf z}{\cal D}^{mn}({\bf y},{\bf z})
{\delta A_{{\sf GI}i}^l({\bf z})\over\delta A_j^c({\bf x})} 
{\partial\over\partial z_i}\delta({\bf z}-{\bf y})\nonumber \\
&&=-gf^{lnm}R_{bc}({\bf x})
\int d{\bf y}\int d{\bf z}{\cal D}^{mn}({\bf y},{\bf z})
\left[\Pi_{j}^c({\bf x}),A_{{\sf GI}i}^l({\bf z})\right] 
{\partial\over\partial z_i}\delta({\bf z}-{\bf y})\label{eq:B2}
\eea
where we have used Eq.~(\ref{eq:reneq1}).
Substituting Eq.~(\ref{eq:PAcomm}) and using
\be
\Pi_{j}^c({\bf x})=R_{bc}({\bf x})\Pi_{{\sf GI}j}^b({\bf x})\,,
\ee 
we obtain that
\be
I_j^b({\bf x})=igf^{bca}{\partial\over\partial x_j}{\cal D}^{ac}({\bf x},{\bf x})
+{\partial\over\partial x_j}\phi^b({\bf x})
\label{eq:B4}
\ee
where the gradient acts only on the first argument 
of ${\cal D}^{ac}$ and the longitudinal term comes from the second term of 
Eq.~(\ref{eq:PAcomm}),
\be
\phi^b({\bf x})=igf^{bca}\int d{\bf y}{\partial\over\partial y_j}{\cal D}^{ac}({\bf y},{\bf y})
{\cal D}^{mn}({\bf x},{\bf y})\stackrel{\Longleftarrow}{D_j^{nm}}({\bf y})
\label{eq:B5}
\ee 
with ${\partial\over\partial y_i}$ acting on the first argument of ${\cal D}^{ac}({\bf y},{\bf y})$.
Comparing the transverse part of Eq.~(\ref{eq:B4}) with that of Eq.~(\ref{eq:Tcomm}),  Eq.~(\ref{eq:reneq2}) is proved.

\section{}
\label{sec:Appc}
To prove the hermiticity of the Faddeev-Popov determinant ${\cal J}$ as an operator in the 
Hilbert space of states,
we recall the criterion that an operator is hermitian if its expectation 
values with respect to all states are real. In the coordinate 
representation of states for which ${\sf A}_{\sf GI}$ is diagonalized and 
corresponds to a c-number field configuration, the 
expectation value of an operator which is a functional of the operator 
${\sf A}_{\sf GI}$ is equal to the same functional of the c-number field 
configuration ${\sf A}_{\sf GI}$. For each c-number field configuration, the 
Faddeev-Popov operator, ${\bf \partial}_jD_j$, with $D_j$ 
denoting the covariant derivative, $D_j^{ab}=\delta^{ab}\partial_j-gf^{abc}A_{{\sf GI}j}^c$,
becomes an operator with respect to space coordinates 
and group indices. We have
\be
\partial_j^\dagger=-\partial_j\label{eq:C1}
\ee
and
\be
D_j^\dagger=-D_j,\label{eq:C2}
\ee
with the dagger referring to space coordinates and group indices. Therefore
\be
\left(\partial_jD_j\right)^\dagger=D_j\partial_j
=\partial_jD_j,\label{eq:C3}
\ee
where the last step follows from the transversality of $A_{\sf GI}$. Therefore 
the Faddeev-Popov operator is hermitian with space coordinates and group
indices for any field configuration. Its determinant, ${\cal J}$ must be 
real. The hermiticity of ${\cal J}$ in the Hilbert space of states is established according to our
criterion, and the hermiticity of ${\cal J}^{\frac{1}{2}}$ is an obvious corollary.\s

To derive the commutation relations among ${\cal P}_j^l({\bf x})$'s and 
$A_{{\sf GI}j}^l({\bf x})$'s, we notice that
\be
{\cal P}_j^l({\bf x})=\Pi_{{\sf GI}j}^{l{\sf T}}({\bf x})
+{1\over 2}\left[\Pi_{{\sf GI}j}^{l{\sf T}}({\bf x}),\ln{\cal J}\right]\label{eq:C4}
\ee
with the second term a functional of $A_{{\sf GI}j}^l({\bf x})$ only. Then we have
\be
\left[{\cal P}_i^a({\bf x}),A_{{\sf GI}j}^b({\bf y})\right]
=\left[\Pi_{{\sf GI}i}^{a{\sf T}}({\bf x}),A_{{\sf GI}j}^b({\bf y})\right].\label{eq:C5}
\ee
Furthermore
\bea 
&&\left[{\cal P}_i^a({\bf x}),{\cal P}_j^b({\bf y})\right]
=\left[\Pi_{{\sf GI}i}^{a{\sf T}}({\bf x}),\Pi_{{\sf GI}j}^{b{\sf T}}({\bf y})\right]\nonumber \\
&&+{1\over 2}\left[\Pi_{{\sf GI}i}^{a{\sf T}}({\bf x}),
\left[\Pi_{{\sf GI}j}^{b{\sf T}}({\bf y}),\ln{\cal J}\right]\right]
+{1\over 2}\left[\left[\Pi_{{\sf GI}i}^{a{\sf T}}({\bf x}),\ln{\cal J}\right],
\Pi_{{\sf GI}j}^{b{\sf T}}({\bf y})\right]\nonumber \\
&&=\left[\Pi_{{\sf GI}i}^{a{\sf T}}({\bf x}),\Pi_{{\sf GI}j}^{b{\sf T}}({\bf y})\right]=0,\label{eq:C6}
\eea
where the Jacobian identity
\bea
&&\left[\Pi_{{\sf GI}i}^{a{\sf T}}({\bf x}),
\left[\Pi_{{\sf GI}j}^{b{\sf T}}({\bf y}),\ln{\cal J}\right]\right]
+\left[\left[\Pi_{{\sf GI}i}^{a{\sf T}}({\bf x}),\ln{\cal J}\right],
\Pi_{{\sf GI}j}^{b{\sf T}}({\bf y})\right]\nonumber \\
&&=-\left[\ln{\cal J},\left[\Pi_{{\sf GI}i}^{a{\sf T}}({\bf x}),
\Pi_{{\sf GI}j}^{b{\sf T}}({\bf y})\right]\right]=0.\label{eq:C7}
\eea
has been employed.
Therefore the commutation relations among ${\cal P}_j^l({\bf x})$'s and 
$A_{{\sf GI}j}^l({\bf x})$'s remain canonical.

\section{}
\label{sec:Appd}
In this section, we shall prove Eq.~(\ref{eq:rb}). Using Eq.~(\ref{eq:psic}), we define 
\be 
\Psi_n^{\sf T}=\int d{\bf r_1}...d{\bf r_n}\overline{{\cal A}_{j_1}^{b_1{\sf T}}}({\bf r}_1)...
\overline{{\cal A}_{j_n}^{b_n{\sf T}}}({\bf r}_n)\Pi_{j_1}^{b_1{\sf T}}({\bf r}_1)...\Pi_{j_n}^{b_n{\sf T}}
({\bf r}_n)\,,\label{eq:D5}
\ee
from which it follows that 
\be
\left[\Pi_i^a({\bf x}),\Psi_n^{\sf T}\right]=n\int d{\bf y}\left[\Pi_i^a({\bf x}),
\overline{{\cal A}_{j}^{b{\sf T}}}
({\bf y})\right]\Psi_{n-1}^{\sf T}\Pi_j^b({\bf y}).\label{eq:D6}
\ee
This leads to
\be
\Pi_i^a({\bf x})\Psi^{\sf T}=\Psi^{\sf T}\Pi_i^a({\bf x})+i\int d{\bf y}
\left[\Pi_i^a({\bf x}),\overline{{\cal A}_{j}^{b{\sf T}}}({\bf y})\right]\Psi^{\sf T}\Pi_j^b({\bf y}).
\label{eq:D7}
\ee

The commutator involved can be calculated from the relation
\be
\overline{{\cal A}_{j}^{b\,{\sf T}}}({\bf x})=A_{{\sf GI}j}^b({\bf x})-A_{j}^{b{\sf T}}({\bf x}),\label{eq:D8}
\ee
which implies that
\be
\left[\Pi_i^a({\bf x}),\overline{{\cal A}_{j}^{b\,{\sf T}}}({\bf y})\right]=
-i{\delta A_{{\sf GI}j}^b({\bf y})\over\delta A_i^a({\bf x})}
+i\delta_{ab}\delta_{ij}^{\sf T}({\bf x}-{\bf y}).
\label
{eq:D9}
\ee
The functional derivative was calculated in Ref. ~\cite{christlee} and the commutator
in Ref.~\cite{HGrib}, and the result  can also be deduced from Eq.~(\ref{eq:PAcomm}) with 
the aid of Eq.~(\ref{eq:Pigi2}), which gives rise to
\be
\left[\Pi^{a}_{i}({\bf x})\,,A_{{\sf GI}j}^{b}({\bf{y}})\right]=
-i\left(R_{ba}({\bf{x}})\delta_{ij}
\delta({\bf x}-{\bf y})+R_{la}({\bf{x}}){\partial\over {\partial}x_i}
{\cal D}^{\,lk}({\bf x},{\bf y})\stackrel{\Longleftarrow}{
D^{kb}_{j}}({\bf y})\right)
\label{eq:D10}
\ee
Substituting this into Eq.~(\ref{eq:D7}) and using Eq.~(\ref{eq:Pigi2}),
we find that
\be
\Pi_{{\sf GI}i}^a({\bf x})\Psi|\phi>=
\Psi\Pi_{i}^{a{\sf T}}({\bf x})|\phi>-{\partial\over\partial x_i}\int d{\bf y}
{\cal D}^{\,al}({\bf x},{\bf y})\stackrel{\Longleftarrow}{
D^{lb}_{j}}({\bf{y}})\Psi\Pi_{j}^{b{\sf T}}({\bf y})|\phi>.\label{eq:D12}
\ee
Taking the transverse part of both side, we end up with
\be
\Pi_{{\sf GI}i}^{a{\sf T}}({\bf x})\Psi|\phi>=\Psi\Pi_{i}^{a{\sf T}}({\bf x})|\phi>.\label{eq:D13}
\ee
The identity Eq.~(\ref{eq:rb}) is proved.
\section{}
\label{sec:appe0}

In this Appendix, we will show how to obtain Eq.~(\ref{eq:Hnew}) from Eq.~(\ref{eq:h1}).
In order to obtain the bilinear product ${\cal P}^{b\,{\sf T}}_{j}({\bf r}){\cal P}^{b\,{\sf T}}_{j}({\bf r})$
for inclusion in a non-interacting part of $({\hat { H}}_{\sf GI})_{\mbox{phys}}$ that can define interaction picture
operators, we now express ${\cal J}^{-\frac{1}{2}}
{\cal P}_{j}^{b\,{\sf T}}({\bf r}){\cal J}{\cal P}^{b\,{\sf T}}_{j}({\bf r}){\cal J}^{-\frac{1}{2}}$ as 
\bea
{\cal J}^{-\frac{1}{2}}{\cal P}_{j}^{b\,{\sf T}}({\bf r}){\cal J}
{\cal P}^{b\,{\sf T}}_{j}({\bf r}){\cal J}^{-\frac{1}{2}}&&=
{\cal P}_{j}^{b\,{\sf T}}({\bf r}){\cal P}_{j}^{b\,{\sf T}}({\bf r})+{\cal P}^{b\,{\sf T}}_{j}({\bf r})
{\cal J}^{\frac{1}{2}}\left[{\cal P}^{b\,{\sf T}}_{j}({\bf r})\,,{\cal J}^{-\frac{1}{2}}\right]-\nonumber\\
&&\!\!\!\!\!\!\!\!\!\!\!\!\!\!\!\!\!\!\!\!\!\!\!\!
\left[{\cal P}^{b\,{\sf T}}_{j}({\bf r})\,,{\cal J}^{-\frac{1}{2}}\right]{\cal J}^{\frac{1}{2}}
{\cal P}^{b\,{\sf T}}_{j}({\bf r})-\left[{\cal P}^{b\,{\sf T}}_{j}({\bf r})\,,{\cal J}^{-\frac{1}{2}}\right]
{\cal J}\left[{\cal P}^{b\,{\sf T}}_{j}({\bf r})\,,{\cal J}^{-\frac{1}{2}}\right]
\label{eq:IPIPI}
\eea 
and make use of the formula
\be
\left[{\cal P}^{b\,{\sf T}}_{j}({\bf r})\,,{\cal J}^{-\frac{1}{2}}\right]=-{\ty \frac{1}{2}}{\cal J}^{-\frac{3}{2}}
\left[{\cal P}^{b\,{\sf T}}_{j}({\bf r})\,,{\cal J}\right]=-{\ty \frac{1}{2}}{\cal J}^{-\frac{1}{2}}
\left[{\cal P}^{b\,{\sf T}}_{j}({\bf r})\,,\ln({\cal J})\right]\,.
\ee
We also observe that the commutator of ${\cal P}_{j}^{b\,{\sf T}}({\bf r})$ and any functional of 
$A^a_{{\sf GI}\,i}({\bf r}^\prime)$ commutes with any other functional of $A^a_{{\sf GI}\,i}({\bf r}^\prime)$, and that,
in fact,
\be
\left[{\cal P}_{j}^{b\,{\sf T}}({\bf y})\,,A_{{\sf GI}\,i}^{a}({\bf{x}})\right]=
\left[\Pi^{b\,{\sf T}}_{{\sf GI}\,j}({\bf y})\,,A_{{\sf GI}\,i}^{a}({\bf{x}})\right]
=-i\delta_{ab}\left(\delta_{ij}-\frac{\partial_i\partial_j}{\partial^2}\right)\delta({\bf x}-{\bf y})\,.
\label{eq:PTcomm}
\ee 
With these observations, we obtain 
\bea
{\cal J}^{-\frac{1}{2}}{\cal P}_{j}^{b\,{\sf T}}({\bf r}){\cal J}
{\cal P}^{b\,{\sf T}}_{j}({\bf r}){\cal J}^{-\frac{1}{2}}&&=
{\cal P}_{j}^{b\,{\sf T}}({\bf r}){\cal P}_{j}^{b\,{\sf T}}({\bf r})-{\ty \frac{1}{2}}
\left[{\cal P}^{b\,{\sf T}}_{j}({\bf r})\,,\left[{\cal P}^{b\,{\sf T}}_{j}({\bf r})\,,\ln({\cal
J})\right]\,\right]\nonumber\\ &&\!\!\!\!\!\!\!\!\!
-{\ty \frac{1}{4}}\left[{\cal P}^{b\,{\sf T}}_{j}({\bf r})\,,
\ln({\cal J})\right] \left[{\cal P}^{b\,{\sf T}}_{j}({\bf r})\,,
\ln({\cal J})\right]\,.
\label{eq:IPIPIa}
\eea
Eqs.~(\ref{eq:reneq2}) and (\ref{eq:Tcomm}) show that 
\be
\left[{\cal P}^{b\,{\sf T}}_{j}({\bf y})\,,\ln({\cal J})\right]=
\left[\Pi_{{\sf GI}\,j}^{b\,{\sf T}}({\bf y})\,,\ln({\cal J})\right]=
igf^{hcb}\delta_{j,k}^{\sf T}({\bf y})\lim_{{\bf x}\ra{\bf y}}{\frac{\partial}{{\partial}y_k}}{\cal D}^{\,ch}({\bf y},{\bf x})\,.
\label{eq:RPJP}
\ee
With Eq.~(\ref{eq:PIe}), this can be rewritten in the form 
\be
\left[\Pi_{{\sf GI}\,j}^{b\,{\sf T}}({\bf y})\,,\ln({\cal J})\right]=
ig^2f^{bdh}f^{{\delta}ds}\delta_{j,k}^{\sf T}({\bf y})
{\int}d{\bf z}{ \frac{\partial}{{\partial}y_k}} 
\left(\frac{1}{4{\pi}|{\bf y}-{\bf z}|}\right)A_{{\sf
GI}\,l}^{{\delta}} ({\bf z})\frac{\partial}{{\partial}z_l}{\cal D}^{\,sh}({\bf z},{\bf y})
\label{eq:RPJPb}
\ee
where $\delta_{j,k}^{\sf T}({\bf y})=\left(\delta_{jk}-
{ \frac{\partial^{({\bf y})}_j\partial^{({\bf y})}_k}{\partial^2}}\right)\delta({\bf y})$. In this form, it is 
clear that, to leading order, $\left[\Pi_{{\sf GI}\,j}^{b\,{\sf T}}({\bf y})\,,\ln({\cal J})\right]$ 
is a $g^2$ term and that the limit ${\bf x}\ra{\bf y}$ has already been carried out.
We can use Eqs.~(\ref{eq:IPIPIa}) and (\ref{eq:RPJPb}) to obtain an expression 
for ${\cal U}\equiv-\frac{1}{4}{\int}d{\bf r}\left[{\cal P}^{b\,{\sf T}}_{j}({\bf r})\,,
\ln({\cal J})\right] \left[{\cal P}^{b\,{\sf T}}_{j}({\bf r})\,,
\ln({\cal J})\right]$, which becomes an interaction term in $({\hat { H}}_{\sf GI})_{\mbox{phys}}$, given by 
\bea
{\cal U}=&&{\ty
\frac{1}{8}}\,g^4f^{bdh}f^{{\delta}ds}f^{bd^{\prime}h^{\prime}}f^{{\delta^\prime}d^{\prime}s^{\prime}}{\int}d{\bf
y}d{\bf z} d{\bf z}^{\prime}\left\{\delta_{j,k}^{\sf T}({\bf y})\delta_{j,k^{\prime}}^{\sf T}({\bf y}){
\frac{\partial}{{\partial}y_k}} 
\left(\frac{1}{4{\pi}|{\bf y}-{\bf z}|}\right)A_{{\sf
GI}\,l}^{{\delta}} ({\bf z})\right.\times\nonumber\\
&&\left.\frac{\partial}{{\partial}z_l}{\cal D}^{\,sh}({\bf z},{\bf y}){ \frac{\partial}{{\partial}y_{k^{\prime}}}} 
\left(\frac{1}{4{\pi}|{\bf y}-{\bf z}^{\prime}|}\right)A_{{\sf
GI}\,{l^{\prime}}}^{{\delta}^\prime} ({\bf z}^{\prime})\frac{\partial}{{\partial}z^{\prime}_{l^{\prime}}}
{\cal D}^{\,s^{\prime}h^{\prime}}({\bf z}^{\prime},{\bf y})\right\}\,.
\label{eq:U}
\eea
Similarly, from Eq.~(\ref{eq:RPJP}), we see that
\be
\left[{\cal P}^{b\,{\sf T}}_{j}({\bf y})\,,\left[{\cal P}^{b\,{\sf T}}_{j}({\bf y})\,,\ln({\cal
J})\right]\,\right]=igf^{hdb}\left[{\cal P}^{b\,{\sf T}}_{j}({\bf y})\,,
\delta_{j,k}^{\sf T}({\bf y})\lim_{{\bf x}\ra{\bf y}}
{\frac{\partial}{{\partial}y_k}}{\cal D}^{\,dh}({\bf y},{\bf x})\right]
\label{eq:Ua}
\ee 
in which we represent ${\cal P}^{b\,{\sf T}}_{j}({\bf y})$ as
$\lim_{{\bf r}\ra{\bf y}}{\cal P}^{b\,{\sf T}}_{j}({\bf r})$ so that 
\be
\left[{\cal P}^{b\,{\sf T}}_{j}({\bf y})\,,\left[{\cal P}^{b\,{\sf T}}_{j}({\bf y})\,,\ln({\cal
J})\right]\,\right]=igf^{hdb}\lim_{{\bf r}\ra{\bf y}}\delta_{j,k}^{\sf T}({\bf y})
\lim_{{\bf x}\ra{\bf y}}{\frac{\partial}{{\partial}y_k}}\left[{\cal P}^{b\,{\sf T}}_{j}({\bf r})\,,
{\cal D}^{\,dh}({\bf y},{\bf x})\right]\,.
\label{eq:Ub}
\ee
Using Eq.~(\ref{eq:Pihermg}), we obtain 
\bea
\left[{\cal P}^{b\,{\sf T}}_{j}({\bf r})\,,{\cal D}^{\,dh}({\bf y},{\bf x})\right]
\!\!\!\!\!\!\!\!\!&&=-gf^{s\alpha t}
{\int}d{\bf z}\,{\cal D}^{\,ds}({\bf y},{\bf z})\left[{\cal P}^{b\,{\sf T}}_{j}({\bf
r})\,, A_{{\sf GI}\;\ell}^{\alpha}({\bf z})\right]
\partial^{({\bf z})}_\ell{\cal D}^{\,th}({\bf z},{\bf x})=\nonumber\\
&&\!\!\!\!\!\!\!\!\!\!\!\!\!\!\!-igf^{sbt}
{\int}d{\bf z}\,{\cal D}^{\,ds}({\bf y},{\bf z})\left(\delta_{j,\ell}-
\frac{\partial^{({\bf z})}_j
\partial^{({\bf z})}_\ell}{\partial^2}\right)\delta({\bf z}-{\bf r})
\partial^{({\bf z})}_\ell{\cal D}^{\,th}({\bf z},{\bf x})\,.
\label{eq:Uc}
\eea 
After integration over all of space, 
${\int}d{\bf y}\left[{\cal P}^{b\,{\sf T}}_{j}({\bf y})\,,\left[{\cal P}^{b\,{\sf T}}_{j}({\bf y})\,,\ln({\cal
J})\right]\,\right]$ becomes another interaction term in $({\hat { H}}_{\sf GI})_{\mbox{phys}}$,
given by
\be
{\cal V}={\ty \frac{1}{4}}g^2f^{hdb}f^{sbt}{\int}d{\bf y}\lim_{{\bf r}\ra{\bf y}}
\left\{\delta_{j,k}^{\sf T}({\bf y})\delta_{j,\ell}^{\sf T}({\bf r})\left[\partial^{({\bf y})}_k
{\cal D}^{\,ds}({\bf y},{\bf r})
\partial^{({\bf r})}_\ell{\cal D}^{\,th}({\bf r},{\bf y})\right]\right\}.
\label{eq:Ud}
\ee
${\cal V}$ is singular since the leading terms in ${\cal D}^{\,ds}({\bf y},{\bf r})$ and 
${\cal D}^{\,th}({\bf r},{\bf y})$,
($-\delta_{ds}\left(4{\pi}|{\bf y}-{\bf r}|\right)^{-1}$ and $-\delta_{th}\left(4{\pi}|{\bf y}-{\bf r}|\right)^{-1}$
respectively), are not eliminated by the structure constants in ${\cal V}$. Christ and Lee called attention to such 
singularities in their work,~\cite{christlee} and conjectured that they might be useful in cancelling unresolved 
divergences in Coulomb-gauge QCD. The same remark applies to ${\cal V}$.
We continue by eliminating the nonhermitian chromoelectric fields from $J_{0\,({\sf GI})}^{a\,{\sf T}}$ and 
$J_{0\,({\sf GI})}^{a\,{\sf T}\,\dagger}$, obtaining\be
J_{0\,({\sf GI})}^{c\,{\sf T}}({\bf y})=gf^{cqp}A^q_{{\sf GI}\,j}({\bf y})\Pi_{{\sf GI}\,j}^{p\,{\sf T}}({\bf y})=
gf^{cqp}A^q_{{\sf GI}\,j}({\bf y})\left({\cal P}_{j}^{p\,{\sf T}}({\bf y})-{\ty \frac{1}{2}
\left[{\cal P}_{j}^{p\,{\sf T}}({\bf y})\,,\ln({\cal J})\right]}\right)
\label{eq:JjPa1}
\ee
and
\be
J_{0\,({\sf GI})}^{c\,{\sf T}\,\dagger}({\bf y})=gf^{cqp}A^q_{{\sf GI}\,j}({\bf y})
\Pi_{{\sf GI}\,j}^{p\,{\sf T\,\dagger}}({\bf y})=
gf^{cqp}A^q_{{\sf GI}\,j}({\bf y})\left({\cal P}_{j}^{p\,{\sf T}}({\bf y})+{\ty \frac{1}{2}
\left[{\cal P}_{j}^{p\,{\sf T}}({\bf y})\,,\ln({\cal J})\right]}\right)\,,
\label{eq:JjPb1}
\ee 
and
\be
J_{0\,({\sf GI})}^{c\,{\sf T}}({\bf y})={\bar {\sf J}}_{0\,({\sf GI})}^{c\,{\sf T}}({\bf y})+i{\sf k}^c_0({\bf y})
\;\;\mbox{and}\;\;J_{0\,({\sf GI})}^{c\,{\sf T\,\dagger}}({\bf y})={\bar {\sf J}}_{0\,({\sf GI})}^{c\,{\sf T}}
({\bf y})-i{\sf k}^c_0({\bf y})
\label{eq:JjPa2}
\ee
where 
\be
{\bar {\sf J}}_{0\,({\sf GI})}^{c\,{\sf T}}=gf^{cqp}A^q_{{\sf GI}\,j}{\cal P}_{j}^{p\,{\sf T}}
\label{eq:Jbar}
\ee 
and,
using Eq.~(\ref{eq:PIe}), $i{\sf k}^c_0$ can be identified as an additional, auxiliary gluon color-charge density in which
\be
{\sf k}^c_0({\bf y})=-{\ty \frac{1}{2}}g^3f^{cqp}f^{hdp}f^{{\gamma}ds}A^q_{{\sf GI}\,i}({\bf y})
\delta_{i,j}^{\sf T}({\bf y}){\int}d{\bf z}{ \frac{\partial}{{\partial}y_j}} 
\left(\frac{1}{4{\pi}|{\bf y}-{\bf z}|}\right)A_{{\sf GI}\,k}^{{\gamma}} 
({\bf z})\frac{\partial}{{\partial}z_k}{\cal D}^{\,sh}({\bf z},{\bf y})\,.
\label{eq:JjPa3}
\ee
This representation enables us to express ${ H}_{\sf C}$ --- the nonlocal interaction involving quark and 
gluon color-charge densities in Eq.~(\ref{eq:h1}) --- in the manifestly hermitian form 
\be 
\!\!\!\!\!\!{ H}_{\sf C}=-{\ty \frac{1}{2}}\int\!d{\bf r}d{\bf x}d{\bf y}\left(j_0^b({\bf x})+
{\bar {\sf J}}_{0\,({\sf GI})}^{b\,{\sf T}}({\bf x})-i{\sf k}^b_0({\bf x})\right)
\stackrel{\longleftarrow}{{\cal D}^{\,ab}}({\bf r},{\bf x}){\partial^2}
{\cal D}^{\,ac}({\bf r},{\bf y})\left(j_0^c({\bf y})+
{\bar {\sf J}}_{0\,({\sf GI})}^{c\,{\sf T}}({\bf y})+i{\sf k}^c_0({\bf y})\right)
\ee
in which all operator-valued fields are hermitian as well as gauge-invariant.
When we have eliminated all the $\Pi_{{\sf GI}\,j}^{p\,{\sf T}}$ and $\Pi_{{\sf GI}\,j}^{p\,{\sf T\,\dagger}}$
from $({\hat { H}}_{\sf GI})_{\mbox{phys}}$ and replaced them with ${\cal P}_{j}^{p\,{\sf T}}$ and the 
other expressions obtained in this process, we obtain Eq.~(\ref{eq:Hnew}).
\section{}
\label{sec:Appf}
To establish Eq.~(\ref{eq:final1}), we quote an identity in \cite{christlee}, 
\be
f^{abc}\int d{\bf r}\Big[D_j^{am}X^m({\bf r})Y^b({\bf r})Z^c({\bf r})
+X^a({\bf r})D_j^{bm}Y^m({\bf r})Z^c({\bf r})+X^a({\bf r})Y^b({\bf r})D_j^{cm}Z^m({\bf r})\Big]=0.
\label{eq:E1}
\ee
The proof follows from the observation that the ordinary derivative terms of the covariant derivatives, 
$D_j$'s, in Eq.~(\ref{eq:E1}) add up to a total derivative 
and the structure constant terms of $D_j$'s add up to zero on account of the Jacobian identity
\be
f^{lab}f^{lmc}+f^{lbc}f^{lma}+f^{lca}f^{lmb}=0.
\label{eq:E2}
\ee
Notice that, the functions $X$, $Y$ and $Z$ may carry other color or vector indices and the dependence on 
other coordinates.\s

According to Eq.~(\ref{eq:Uc}), 
\be
\Big[\Pi_j^c({\bf r}),{\cal D}^{ab}({\bf x},{\bf y})\Big]=-\int d{\bf z}d{\bf z}^\prime
{\cal D}^{am}({\bf x},{\bf z})\left[\Pi_j^c({\bf x}),(\partial\cdot D)^{mn}\right]\delta({\bf z}-{\bf z}^\prime)
{\cal D}^{nb}({\bf z}^\prime,{\bf y})
\label{eq:E3}
\ee
and, with Eq.~(\ref{eq:D10}), we have
\bea
&&\Big[E_j^b({\bf r}),\Delta_j^b({\bf r})\Big]=-\Big[\Pi_{{\sf GI},j}^b({\bf r}),\Delta_j^b({\bf r})\Big]
=-R_{ba}({\bf r})\Big[\Pi_j^a({\bf r}),\Delta_j^b({\bf r})\Big]\nonumber \\
&&={1\over 2}g^2f^{lab}f^{lmn}\partial_j{\cal D}^{am}({\bf r},{\bf r})\partial_j{\cal D}^{nb}({\bf r},{\bf r})
+{g^2\over 2}f^{lab}f^{cmn}\int d{\bf x}D_i^{ck}{\cal D}^{kl}({\bf x},{\bf r})\stackrel{\leftarrow}
{\partial}_j{\cal D}^{ma}({\bf x},{\bf r})\stackrel{\leftarrow}{\partial}_j
\partial_i{\cal D}^{nb}({\bf x},{\bf r}),
\label{eq:E4}
\eea
where the symmetry property Eq.~(\ref{eq:symmetry}) is employed to obtain the second term on the right hand side.
Upon relabeling the dummy color indices, we have
\bea
&&{\hbox{The 2nd term of r. h. s. of Eq.~(\ref{eq:E4})}}={g^2\over 4}f^{lab}f^{cmn}\int d{\bf x}
D_i^{ck}{\cal D}^{kl}({\bf x},{\bf r})\stackrel{\leftarrow}
{\partial}_j{\cal D}^{ma}({\bf x},{\bf r})\stackrel{\leftarrow}{\partial}_j
\partial_i{\cal D}^{nb}({\bf x},{\bf r})\nonumber \\
&&+{g^2\over 4}f^{lab}f^{cmn}\int d{\bf x}{\cal D}^{cl}({\bf x},{\bf r})\stackrel{\leftarrow}
{\partial}_jD_i^{mk}{\cal D}^{ka}({\bf x},{\bf r})\stackrel{\leftarrow}{\partial}_j
\partial_i{\cal D}^{nb}({\bf x},{\bf r})\nonumber \\
&&=-{g^2\over 4}f^{lab}f^{cmb}\partial_j{\cal D}^{lc}({\bf r},{\bf r})\partial_j{\cal D}^{am}({\bf r},{\bf r}),
\label{eq:E5}
\eea
where the last step follows from the identities Eqs.~(\ref{eq:E1}) and (\ref{eq:FPd}). We have, then 
\be
\left[E_j^b({\bf r}),\Delta_j^b({\bf r})\right]={1\over 4}g^2(2f^{lbm}f^{lan}-f^{lam}f^{lbn})
\partial_j{\cal D}^{ab}({\bf r},{\bf r})\partial_j{\cal D}^{mn}({\bf r},{\bf r})\,.
\label{eq:E6}
\ee
Combining it with 
\be
-\Delta_j^b({\bf r})\Delta_j^b({\bf r})={1\over 4}g^2f^{lab}f^{lmn}
\partial_j{\cal D}^{ab}({\bf r},{\bf r})\partial_j{\cal D}^{mn}({\bf r},{\bf r})
\label{eq:E7}
\ee
according to Eq.~(\ref{eq:integrand}) and using the Jacobian identity Eq.~(\ref{eq:E2}), we end up with
Eq.~(\ref{eq:final1}).


\begin{references}
\bibitem{f1}In this work we use nonrelativistic notation, in which all space-time indices 
are subscripted and designate contravariant components of contravariant quantities such as $A_i^a$ or $j_i^a$, 
and covariant components of covariant quantities such as $\partial_i$.
\bibitem{f2}The transformation of the quark field from the ${\cal C}$ to the ${\cal N}$ representation, 
${\cal U_C}^{-1}\psi\,{\cal U_C}=V_{\cal{C}}^{-1}\,\psi\,,$ and the transformation that leads to the gauge-invariant 
quark field in the ${\cal C}$ representation, ${\psi}_{\sf GI}=V_{\cal{C}}\,\psi$, transform in opposite directions. 
In the former case we transform the quark field that is represented by $\psi$ in the ${\cal C}$ representation
to the ${\cal N}$ representation; in the latter case, we transform the quark field that is gauge-invariant 
in the ${\cal N}$ representation --- namely $\psi$ --- to its corresponding ${\cal C}$ representation 
form. The transformations therefore proceed in opposite directions in the two cases.
\end{references}
\end{document}